\PassOptionsToPackage{table}{xcolor}
\PassOptionsToPackage{numbers,sort&compress}{natbib} 

\documentclass[unnumsec,webpdf,modern,medium]{oup-authoring-template}%

\onecolumn 

\graphicspath{{Fig/}}
\newcommand{\societylogo}{}

\theoremstyle{thmstyleone}%
\theoremstyle{thmstyletwo}%
\theoremstyle{thmstylethree}%

\usepackage{graphicx}
\usepackage{hyperref}
\hypersetup{hidelinks}
\newcommand{\sectionref}[1]{\hyperref[#1]{`\nameref*{#1}'}}
\usepackage{etoolbox}
\newcommand{\publicationcopyright}{Published by Oxford University Press. This is an Open Access article distributed under the terms of the Creative Commons Attribution-NonCommercial License (\url{https://creativecommons.org/licenses/by-nc/4.0/}).}
\makeatletter
\newcommand{\patchpublicationcopyright}{%
  \patchcmd{\ps@opening}
    {Copyright and licence statements to be updated by the publisher during production.}
    {\publicationcopyright}
    {}{}%
}
\patchpublicationcopyright
\patchpublicationcopyright
\patchpublicationcopyright
\patchpublicationcopyright
\patchpublicationcopyright
\patchpublicationcopyright
\patchcmd{\@maketitle}
  {\if@modern\if@mediumone\raggedright}
  {\parindent=0pt\relax\if@modern\if@mediumone\raggedright}
  {}{\PackageError{sme-review}{Could not patch abstract indentation}{Check the OUP class abstract layout.}}
\patchcmd{\@maketitle}
  {\if@medium\rightskip-12pt\else\fi}
  {\if@medium\rightskip0pt\else\fi}
  {}{\PackageError{sme-review}{Could not patch abstract right margin}{Check the OUP class abstract layout.}}
\makeatother
\usepackage{array}
\usepackage[section]{placeins}
\usepackage{longtable}

\usepackage{csquotes}
\MakeOuterQuote{"}

\begin{document}

\journaltitle{Journal of Cybersecurity}
\DOI{10.1093/cybsec/tyag030}
\copyrightyear{2026}
\pubyear{2026}
\vol{12}
\volume{12}
\issue{1}
\access{Open Access}

\firstpage{1}


\title[A Systematic Review of SME Cybersecurity]{Unaware, unfunded, untrained and unsupported: a systematic review of cybersecurity in small and medium enterprises}

\author[1,$\ast$]{Carlos Rombaldo Junior\ORCID{0009-0004-0491-3146}}
\author[1]{Ingolf Becker\ORCID{0000-0002-3963-4743}}
\author[1]{Shane D Johnson\ORCID{0000-0002-0184-9896}}


\address[1]{\orgdiv{Department of Security \& Crime Science}, \orgname{University College of London}, \orgaddress{\street{Gower Street}, \postcode{WC1E 6BT}, \state{London}, \country{United Kingdom}}}
\corresp[$\ast$]{Corresponding author. Department of Security \& Crime Science, University College of London, Gower Street, WC1E 6BT London, UK. E-mail: \href{mailto:jr.rombaldo@gmail.com}{jr.rombaldo@gmail.com}}

\received{25}{5}{2025}
\revised{29}{6}{2026}
\accepted{3}{9}{2026}
\makeatletter
\g@addto@macro\@history{. \textbf{Published:} 10 October 2026}
\makeatother





\abstract{
Small and Medium Enterprises (SMEs) are pivotal in the global economy, accounting for over 90\% of businesses and 60\% of employment worldwide. Despite their significance, SMEs are often disregarded in cybersecurity initiatives, rendering them ill-equipped to deal with the growing frequency, sophistication, and destructiveness of cyberattacks.
We systematically reviewed the cybersecurity literature on SMEs published between 2017 and 2024. We focused on research discussing cyber threats, adopted controls, challenges, and constraints SMEs face in pursuing cybersecurity resilience.
Our search yielded 1090 studies that we narrowed to 132 relevant papers. We identified 44 unique themes and categorised them as novel findings or established knowledge. 
This distinction revealed that research on SMEs is underdeveloped and has made little progress in understanding SMEs' roles, threats, and needs. Studies often repeated early discoveries without replicating or offering new insights.
Existing research indicates that the main challenges to attaining cybersecurity resilience of SMEs are a lack of awareness of cybersecurity risks, limited cybersecurity literacy, and constrained financial resources. Resource availability varied between developed and developing countries. Our analysis indicated a relationship among these themes, suggesting that limited literacy is the root cause of awareness and resource constraint issues. 
}


\maketitle

\section{Introduction}
\label{sec:intro}

Services and systems are increasingly digitised and interconnected. While this has clear benefits, the recent boom in online businesses has been associated with an alarming rise in cybercrime~\cite{CyberCrimeEvolution}. In the early 2000s, cyberattacks mostly involved the distribution of viruses and worms with destructive intentions. However, in the past decade, attacks have evolved into large-scale data breaches and ransomware campaigns motivated by financial gain~\cite{CyberCrimeEvolution, Verizon2022}. More recently, we have observed a surge in sophisticated cyber threats from organised crime groups and nation-states~\cite{Huaman2021, Osborn2018, Lynch2020, Cleveland2018}, making it more challenging for organisations to defend against such resourceful adversaries. It is also commonly believed that cyberattacks have increased in scale and sophistication, and attackers have shifted their attention from well-protected large organisations to Small and Medium Enterprises (SME): organisations that typically have fewer resources and less resilient cybersecurity defences~\cite{Pickering2021, Alahmari2020, VanHaastrecht2021, McLilly2020, Heidt2019}.

In this paper, we present the findings from a systematic review (SR) of the growing literature concerned with SME cybersecurity. This review synthesises our understanding of the threats and defences, identifies gaps in knowledge, and provides insights on how to strengthen cybersecurity measures for these enterprises.

Considering the scale of the problem, a common finding reported in the literature is that there has been an increase in the number of successful cyberattacks against SMEs~\cite{Alahmari2020, Johannsen2020, VanHaastrecht2021, Ahmed2021}. For example, according to a Verizon report, 58\% of the attacks registered in 2019 targeted SMEs~\cite{Kabanda2018}, and SMEs were the victims of 43\% of breaches. Furthermore, 60\% of SMEs that suffered a cyberattack go out of business within 6 months~\cite{VerizonSME}. The UK Federation of Small Businesses reported a daily rate of 10\,000 attacks against SMEs based in the UK~\cite{FSB10k}. While these numbers are unsettling, the reality is likely to be worse as SMEs typically do not report the cyberattacks they suffer~\cite{Alahmari2021, McLilly2020, Ikuero2022}.
Although the number of breaches and financial losses continues to increase, recent surveys suggest that SMEs remain unprepared~\cite{White2020, Alabama2019, Mitrofan2020, Carias2021}. For example, a 2022 cybersecurity survey conducted across UK businesses revealed that only 17\% undertook vulnerability audits~\cite{CSBreachesUK2022}. Research suggests that one reason for the lack of preparedness amongst SMEs is their lack of awareness of the threats faced~\cite{Saban2021, Huaman2021, Ncubukezi2020, Lejaka2019, Lynch2020}. In contrast, some researchers believe that the problem lies in SMEs perceiving the risk of an attack as low and consequently not prioritising cybersecurity~\cite{Elezaj2019, Ahmed2021, Kabanda2018}. Other researchers attribute the problem to insufficient investment in cybersecurity~\cite{Ozkan2019, Kljucnikov2019, Mitrofan2020} and insufficient cybersecurity literacy necessary to establish defensive programmes~\cite{Alahmari2020, Raineri2020}.

Readers might wonder what SMEs are and why it is important to research their cyber resilience. To answer the first question, no globally adopted definition of SMEs was found. Instead, researchers have used various criteria, making comparison troublesome (see section~\ref{disc:def}). An example definition is that provided by the European Commission, which defines SMEs as any enterprise with up to 250 employees and EUR 50 million in revenue~\cite{ECSME}.  Here, we do not employ a specific definition but take an inductive approach (section~\ref{sec:data-extraction}) and include any paper that focuses on SMEs (as perceived by the authors of the paper), regardless of their definition.
As for their importance, according to the World Bank, SMEs play a significant role in global economies, representing 90\% of worldwide business, accounting for over 50\% of global employment and 40\% of average national incomes~\cite{WorldBank}. In the USA, they represented 99.9\% of the country's 32.5 million businesses and accounted for 43.5\% of the country's GDP (Gross Domestic Product) in 2020~\cite{USSME}.
Moreover, the statistical office of the European Union~\cite{EUSMESTAT} reported that in 2022, 99.8\% of all enterprises in the 27 EU states were SMEs, with 90\% being micro businesses (less than 10 employees). SMEs also employed 83 million people, the equivalent of 64\% of total employment in the bloc, and were responsible for over half of regional turnover.

\begin{figure}[!htbp]
    \caption{Number of published studies per year in the past decade concerning `SME Cybersecurity'. Data are extracted from Scopus using variations of the terms "Cybersecurity" and "SME" with the requirement that they appeared within 10 words of each other.}
    \label{fig:pubevolution}
    \centering
    \includegraphics[width=0.65\linewidth]{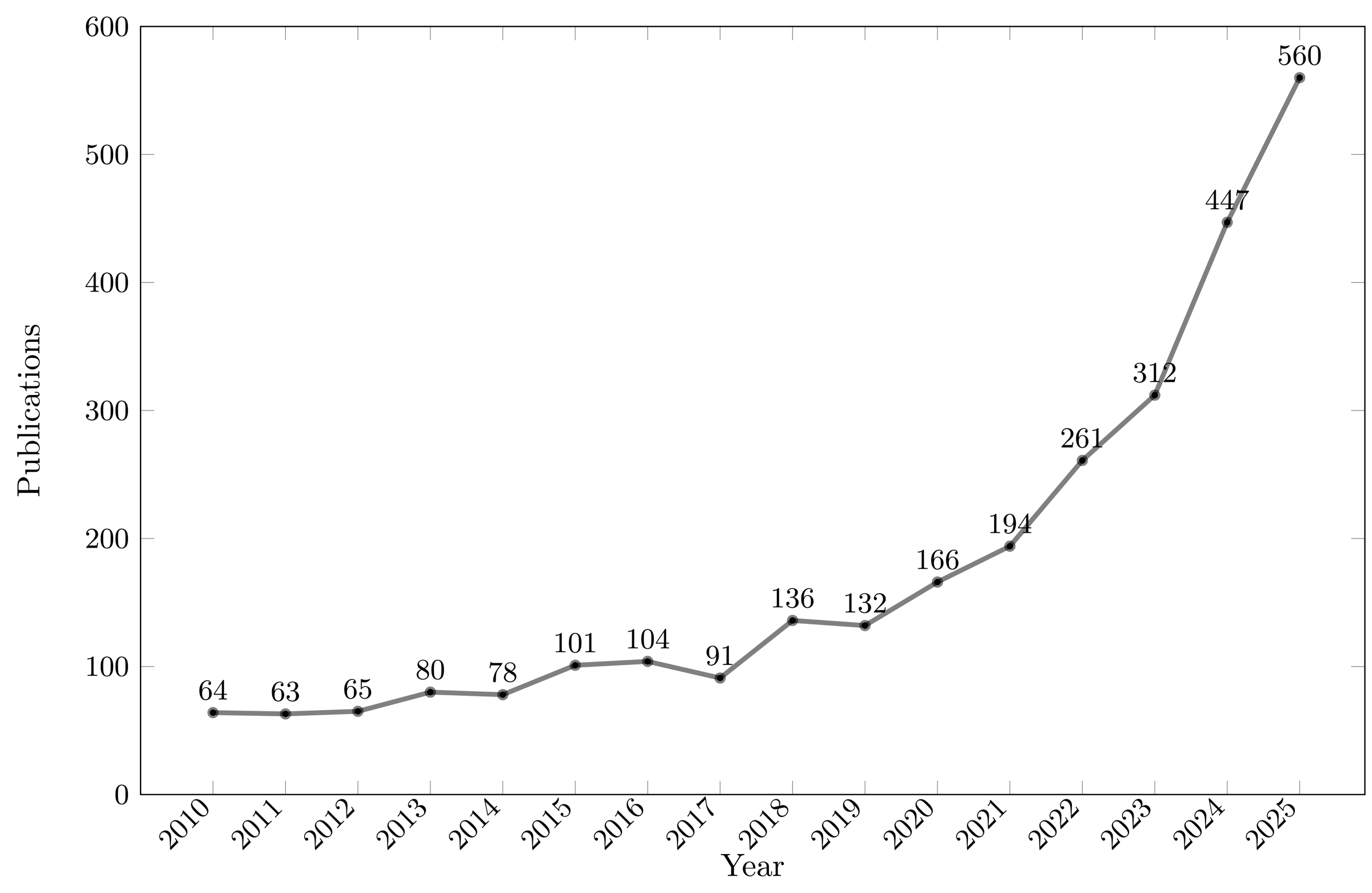}
    \figalttext{Line graph of annual SME-cybersecurity publications from 2010 to 2025, rising from 64 in 2010 to 560 in 2025, with the steepest growth after 2018.}
\end{figure}

For decades, the cybersecurity literature has focused on more prominent (large) businesses, with little attention given to SME-specific threats and how they differ from those directed at larger organisations~\cite{Alabama2019, Heidt2019, Eilts2020, Carias2021}. However, in recent years, the study of SMEs has become a field of inquiry. Figure~\ref{fig:pubevolution} illustrates the rapid and intensified growth of research published on this topic and demonstrates that there has been roughly a 615\% increase in publications over the past 5 years.

Considering the recent surge in research concerning SME cybersecurity, and in the absence of an existing SR on this topic, this paper focuses on synthesising this knowledge through a SR to address the following research questions:

\label{RQ}
\begin{enumerate}
\item \emph{What are the cybersecurity threats experienced by SMEs?}
\item \emph{What awareness do SMEs have about the threats they face?}   
\item \emph{What is the uptake of cybersecurity controls by SMEs?}   
\item \emph{Which existing cybersecurity frameworks apply to or can be tailored to SMEs?}    
\item \emph{What challenges do SMEs face in adhering to existing cybersecurity frameworks and solutions?}
\end{enumerate}

SRs have several advantages over alternative strategies for synthesising existing literature, such as ad-hoc reviews. Conventional review methods can be biased toward studies produced by authors known in the field or aligned with researchers' views (i.e. confirmation bias). SRs address these problems~\cite{Higgins2019, Tranfield2003}.
They rely on a transparent and reproducible protocol designed and agreed upon before data collection begins. This protocol describes the search terms to be used, the data sources to be searched, and the steps involved in extracting and synthesising findings.
As a result, the approach minimises the reviewer's biases in selecting or prioritising studies that confirm their beliefs or interests~\cite{Higgins2019, Braun2008}.
In principle, any two researchers following the same protocol should identify the same studies and arrive at the same conclusions when conducting a SR. 
The primary purpose of a SR is to provide a rigorous, comprehensive, and reproducible consolidation of the state of knowledge on a given topic~\cite{Higgins2019, Tranfield2003}, which is achieved by synthesising evidence from multiple studies, evaluating their quality and consistency, identifying areas of convergence and contradiction, critically exposing gaps where discoveries lack adequate empirical support; and assessing the collective strength and reliability of existing evidence, establishing a foundation upon which future research priorities can be defined.
SRs are commonly used in the field of medicine to synthesise experimental evidence on `what works' to address a given problem~\cite{CrimeSRBook}. However, they have also proven effective in synthesising evidence on emerging topics and dealing with a range of different domains~\cite{Blythe2021, Tranfield2003}. 
In the context of cybersecurity, for which there is rapid technological change, SRs can be particularly valuable because they can reveal whether commonly held assumptions are supported by cumulative, as well as contemporary empirical evidence, or whether conclusions are based on a few or dated findings that have been repeatedly reported across studies without rigorous testing in current or recent environments.

With this in mind, in addition to synthesising the evidence in the usual way, in what follows we also explicitly distinguish between studies that empirically test findings and those that perpetuate them without presenting a test of them. This serves as a form of systematic knowledge gap analysis that goes beyond standard SR synthesis.  Unlike an ordinary gap analysis, which relies on the researcher's expertise to identify gaps, which can be somewhat ad-hoc, our approach is based on an explicit analysis of the literature to identify those topics and conclusions that are frequently discussed but for which new empirical evidence is not presented. 
Section~\ref{sec:gap-analysis} details this technique.

In summary, this paper makes (at least) three contributions. First, it provides a systematic synthesis of the cybersecurity literature concerning SMEs, offering a transparent and reproducible account of what is currently known. Second, through the use of a systematic gap analysis, it reveals a peculiar asymmetry in the literature: many widely cited findings --- such as the notion that cyberattacks against SMEs are rising --- have been propagated across dozens of studies without any of them providing new empirical evidence to support or refute these claims. Third, the review identifies key themes regarding barriers to the development of cyber resilience amongst SMEs.  In doing so, the review aims to inform directions for future research and policy.

The remainder of the paper is organised as follows. The next section describes the protocol employed, including the search engines and search terms used.  This is followed by a synthesis of the findings from the papers reviewed. The final section discusses the implications of the findings, further research opportunities and the limitations of the study.

\section{Methodology}
\label{sec:method}

\begin{figure*}[!htbp]
\centering
 \includegraphics[width=1.0\textwidth]{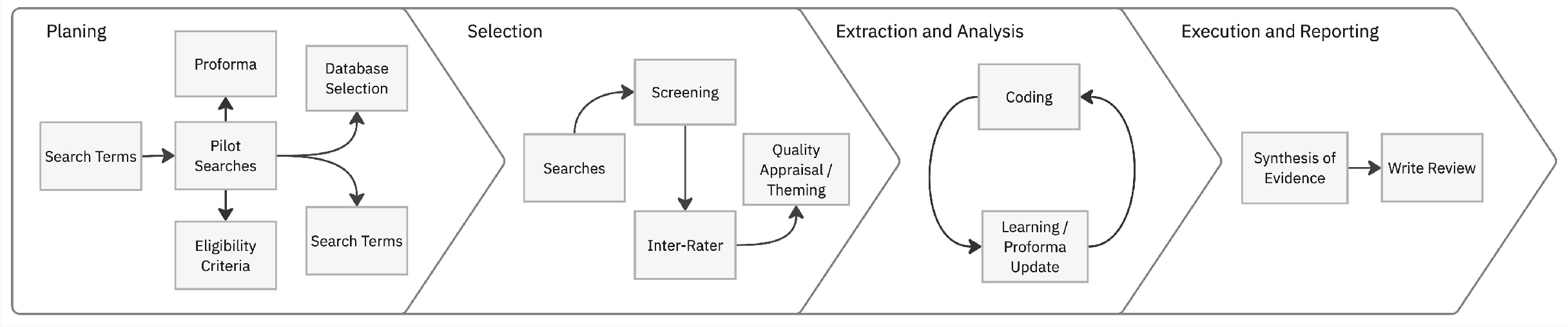}
 \caption{Research method overview.}
 \label{fig:research-method}
 \figalttext{Horizontal flowchart of the review process across four chevron-shaped stages: Planning (search terms, pilot searches, proforma, database selection, eligibility criteria), Selection (searches, screening, inter-rater check, quality appraisal and theming), Extraction and Analysis (iterative loop between coding and proforma update), and Execution and Reporting (synthesis of evidence, write review).}
\end{figure*}

This section discusses the SR methodology employed. The method adheres to PRISMA-P~\cite{Moher2016} and Cochrane~\cite{Higgins2019} frameworks, which are well-established standards for the conduct of SRs. We begin with the elaboration of the research questions (see introduction) using the PICO (Problem, Intervention, Comparison and Outcome) format~\cite{PICO}. We then describe the electronic databases searched, the search terms used, and the eligibility criteria employed to inform the selection of studies. Next, we detail the approaches to data extraction and evidence synthesis. Figure~\ref{fig:research-method} provides an overview of the research method, while the subsections expand on each step.

\subsection{Data sources}

We searched for literature in computer science-focused search engines (ACM Digital and IEEE Xplore Digital) and more general academic search engines (ProQuest, Scopus, and Web of Science). To minimise the danger of publication bias, we also used Google Scholar to search for grey literature and research produced outside of traditional publishing and distribution channels. Examples of grey literature include industry reports, working papers, newsletters, government documents, and white papers~\cite{GreyLit}.  Grey literature was subject to the same inclusion criteria as the academic literature.  Table~\ref{databases-tb} provides further details on each search engine used.

\begin{table*}[!htbp]
    \caption{Databases used to source studies.\label{databases-tb}}
     \small
    \begin{tabular*} {\textwidth} 
    {| p{0.12\linewidth} | p{0.835\linewidth} |}
      \hline
      \multirow{2}{=}{\raggedright Specific academic databases}
      & \textbf{ACM Digital Library}: a comprehensive database of full-text articles and bibliographic literature covering computing and information technology from Association for Computing Machinery publications \\
      & \textbf{IEEE Xplore Digital Library}: indexed articles and papers on computer science, electrical engineering, and electronics from the Institute of Electrical and Electronics Engineers (IEEE) and the Institution of Engineering and Technology \\ 
      \hline
      \multirow{3}{=}{\raggedright Generic academic databases}
      & \textbf{ProQuest}: Databases covered: Library \& Information Science Abstracts (LISA), ProQuest Central (Criminal Justice Database, Computing Database, Library Science Database, Science Database, Social Science Database, Psychology Database and databases covering technology and social sciences, ProQuest Dissertations \& Theses Global\\
      & \textbf{Scopus}: Elsevier's abstract and citation database - Content on Scopus comes from over 5,000 publishers and must be reviewed and selected by an independent Content Selection and Advisory Board (CSAB) to be, and continue to be, indexed on Scopus\\
      & \textbf{Web of Science}: Conference Proceedings Citation Index, Science Citation Index Expanded, Social Sciences Citation Index, Arts \& Humanities Citation Index, and Book Citation Index \\
      \hline
      Grey Literature & \textbf{Google Scholar}: articles, theses, books, abstracts and court opinions from academic publishers, professional societies, online repositories, universities and other websites. \\
      \hline
    \end{tabular*}
\end{table*}

\subsection{Search strategy}

The search query comprised two parts, both refined through pilot searches. The first was used to identify studies related to cybersecurity, while the second located studies with an emphasis on SMEs. 
The following example illustrates the search query used for IEEE Xplore:

\begin{small}
\begin{verbatim}
    ( 'cyber?security' OR 'information security' OR 
      'digital security' OR 'threat' OR 'cyber?threat' OR
      'cyber resilience' OR 'it security' OR 'data security'  )
    NEAR/8
    (  ('SME' OR 'SMME' OR 'SMB')  OR
       ('small' OR 'medium' or 'micro')  NEAR/2 
       ('business' OR 'enterprise' OR 'organi?ation')  )
\end{verbatim}
\begin{verbatim}
NOTE: Search query used for IEEE Xplore. The operator "?" is a single-character wildcard that allows 
for spelling variations  (e.g. "organisation" or "organizations"),  
and the "NEAR/X" operator specifies that words should be within X words of each other.

\end{verbatim}
\end{small}

The exact search query used had to be modified for each search engine according to the syntax supported by that search engine. For example, the ACM Digital Library offers no support for the operator NEAR, which had to be replaced by the `AND' operator.  Moreover, as is typical with SRs, the query had to be limited to titles and abstracts.  Otherwise, it returned hundreds of thousands of additional but irrelevant studies. Similarly, the ProQuest query leveraged the `Not Full Text' operator to match everything except the full text. The Scopus search engine did not support the operator NEAR. Instead, a similar `PRE' function was used. Web of Science could not parse `?', so the query had to be modified to include variations of each term manually. In the case of Google Scholar, this continues to serve pages of results, even when these are no longer relevant. To address this, we continued to review the search results until we encountered several (sequential) pages of search results for which the findings were consistently irrelevant.  Ultimately, we limited our search to the first fifteen pages of the Google Scholar search results. The combination of these diverse databases and carefully constructed search queries minimised the likelihood that relevant articles were missed.

\subsection{Eligibility criteria}

As is standard practice, the screening phase was divided into two parts. The first involved the assessment of study titles and abstracts, while the second involved a full-text review. The following inclusion criteria were applied to screen articles based on their titles and abstracts:

\begin{enumerate}
    \item To ensure that the literature reviewed was relevant to today's SMEs' technological context, articles had to be published on or after 1 January 2017. This date was selected as it coincided with the growth in cloud adoption~\cite{StatistaCloudUsage, StatistaCLoudSpending} and was not long before the forced digitalisation and "Bring Your Own Device" (BYOD) deployments resulted from the restrictions imposed during the COVID-19 pandemic~\cite{Rawindaran2021, Grondys2021, Feher2020, OCED-SME-DigTrans}. Combined with a surge of publications in 2017 (Fig.~\ref{fig:pubevolution}), and the expectation that research is cumulative, we judged the last seven years to be an adequate time frame for this review.
    \item Only literature published in English was included.
\end{enumerate}

For each study that met the inclusion criteria, the full text was obtained, read and assessed based on its methodological and thematic focus. The full text of articles was obtained either through searching open sources, accessing them through the university library, inter-library loans, or by contacting the study author(s) directly. For the full text review, studies were included if they met the following eligibility criteria:

\begin{enumerate}
    \item If they explicitly discussed SME cybersecurity: several cybersecurity studies mentioned SMEs but did not discuss them in relation to cybersecurity, and hence were excluded.
    \item Studies must have thematic relevance. That is, studies must mention cybersecurity threats, controls or solutions to qualify.

    \item Studies had to present a clear and detailed methodology section that would enable their replication and an objective assessment of the approach taken. Specifically, the methodology had to provide details on the following: research method and its rationale, sampling approach and demographics, data collection and validation strategy and data analysis approach.

\end{enumerate}

With any SR, a potential concern with the application of the inclusion criteria is whether any two independent reviewers would apply the inclusion criteria in the same way.  To assess this,  two researchers independently screened a random sample of 5\% (n=52) of the titles and abstracts~\cite{Ratajczyk2016}, and we calculated the Prevalence Adjusted Bias Adjusted Kappa statistic to measure inter-rater reliability (IRR)~\cite{Byrt1993}.  The calculated IRR coefficient of 0.846 indicated `almost perfect agreement'~\cite{Byrt1993}.  Where differences (n = 4) were found, these were discussed and resolved.

The second issue concerns coder drift, that is, the extent to which the same coder applies the inclusion criteria consistently over time.  This can be an issue when a review takes some time to complete.  To minimise this problem, our approach included reviewing the codes from the previous week and the coded content at the beginning of each week. This way, we minimised the danger of diverging codes' interpretation over time.

\subsection{Data collection}

The titles and abstracts for all identified studies were exported into Research Information Systems format files and subsequently imported into Rayyan~\cite{Rayyan2022}, a specialised platform for systematic literature reviews, where the initial selection was performed. As Google Scholar does not include abstracts when exporting results, its bibliography had to be imported into Mendeley for metadata lookup and then into Rayyan.
None of the database exports contained the full texts. As such, studies filtered in the first screening stage were exported from Rayyan to Zotero~\cite{Zotero}, which facilitated the retrieval of full-text PDFs. Abstract and title screening was managed using Rayyan, while full-text screening was completed using Zotero.

\subsection{Data extraction and synthesis}
\label{sec:data-extraction}

A Proforma was produced from pilot searches and sampled studies to assist with data extraction. It consisted of a predefined set of relevant information to be extracted from each study, as follows:

\begin{enumerate}
    \item Year of publication
    \item Theme (Themes emerged from pilot searches and were updated during the screening process. The themes were: Cybersecurity Awareness, Behaviour or Knowledge Gap; Data Security and Privacy; Cybersecurity Incidents and Response; Information Security Management Frameworks (ISMF); IT Security, including BYOD, Internet of Things (IoT), Cloud, Infrastructure and Network Security; Risk Assessment \& Management; Supply-chain Security; Threat Intelligence and Cyber Resilience).
    \item Study type (case study, SR, empirical, quantitative/qualitative, design proposal, industry report, etc.)
    \item Data collection method (observation, surveys, or experimentation)
    \item Size of sampled data (individuals and organisations)
    \item Coverage (geographic, industry and socio-economic)
    \item Discussion of existing cybersecurity frameworks
    \item What motivated the study
    \item Conclusions and recommendations
    \item Limitations, as stated by the authors
\end{enumerate}

We found a limited number (n=6) of studies that provided quantitative data. Therefore, we opted for a qualitative approach to synthesis, combining elements of content and thematic analysis. Content analysis consists of categorising and classifying data according to its objective meaning and measurable attributes. In contrast, thematic analysis represents a more interpretive approach that emphasises the context of the data, making it a more nuanced approach~\cite{Braun2008, ContentThematic}. 
Qualitative analysis of this type identifies patterns in two ways: inductively (bottom-up) or deductively (top-down). The deductive method codes the data according to a predefined set of themes/codes to answer already-established research questions. In contrast, with an inductive approach, research questions and themes emerge during the coding process. The deductive method provides in-depth knowledge of existing themes, whilst the inductive method offers breadth by allowing themes to emerge during data analysis. 
Here, we applied a mixture of inductive and deductive methodologies, referred to by Braun and Clarke as theoretical thematic analysis~\cite{Braun2008}. To do so, we performed an initial literature search to inform the development of a predefined set of themes. 
During the data extraction process, existing codes evolved from arising patterns, and new codes were added in an iterative style until theoretical saturation was reached (i.e., no new themes emerged). At the end of each week, authors reviewed recently emerged codes against existing ones and merged them when applicable.

The first author conducted the data extraction, which was subsequently reviewed and discussed by all authors. This led to a well-defined codebook with clear, agreed-upon definitions. We do not report inter-rater agreement scores for this analysis, as they are considered inappropriate in thematic analysis~\cite{clarkebraun2021}.

\subsection{Systematic knowledge gap analysis}
\label{sec:gap-analysis}
We noted that some themes that emerged from the literature were frequently discussed to motivate research (e.g. that there has been an increase in the number of cyberattacks against SMEs) but that there was little to no evidence of new research that replicated these findings or that tested their validity (e.g. have cyberattacks against SMEs increased within the last 7 years?).
As a form of gap analysis and to identify any asymmetries in the literature, we coded articles as to whether they discussed a theme solely to introduce or motivate the article or if they empirically investigated and (re)tested a topic, thereby adding to the cumulative body of research on the issue. 
We refer to the former as "drivers" or "established knowledge" and the latter as "novel findings" or "conclusions".
We consider this to be important because understanding where asymmetries exist helps to identify priorities for future research. This distinction is particularly relevant given how technology and our use of it have changed over time. Findings that were valid (say) 5 years ago may no longer hold the same relevance.

\subsection{Included studies}
Searches were conducted in May 2024, and the review was completed in September 2024. The initial searches identified 1090 reports, of which 290 were duplicates, and six were unavailable in English and hence excluded.  It is important to note that these searches returned numbers different from Fig.~\ref{fig:pubevolution} because here we used multiple databases and more comprehensive eligibility criteria.

The screening of titles and abstracts resulted in the exclusion of 372 that did not meet the eligibility criteria. Furthermore, two articles were excluded because no full-text article was available.  For one, the journal's website was unavailable at the time of retrieval, and the article could not be accessed through any other means.  For the other, the paper was presented at a conference for which there was no full-text publication. In both cases, the authors did not respond to requests for a copy of the study. Of the 414 studies that made it to the full-text screening, 289 were excluded because they did not meet the eligibility criteria, either because of their thematic focus or because they did not include an adequate methodology section.  Consequently, 132 studies were included in the final review.  Figure~\ref{fig:prisma} summarises the number of articles involved in each stage of this review.

\begin{figure}[!htbp]
\centering
\resizebox{0.9\textwidth}{!}{ 


\centering

\renewcommand{\baselinestretch}{1}

\small

\usetikzlibrary{positioning,chains}

\fontfamily{phv}

\begin{tikzpicture}[
    node distance=15mm and 10mm,
    start chain=going below,
     mytitle/.style = {
        draw, rectangle, align=center, rounded corners, 
        font=\bfseries, 
        inner sep=2.5ex, outer sep=0pt,
        fill={rgb,255:red,220; green,220; blue,220},
        minimum width={0.85\linewidth}, thick,  on chain},
    mynode/.style = {
        draw, rectangle, align=left, text width=6cm,
        font=\small,
        inner sep=2ex, outer sep=0pt,
        thick, on chain},
    mylabel/.style = {
        draw, rectangle, align=center, rounded corners, 
        font=\bfseries, 
        inner sep=2.2ex, outer sep=0pt,
         fill={rgb,255:red,220; green,220; blue,220},
        minimum height=38mm, thick, on chain },
    every join/.style = arrow,
    arrow/.style = {very thick,-stealth}
] 

\coordinate (tc);

\node(n1)[mytitle, right=-35mm of tc] {Identification of studies via databases };

\node (n2)  [mynode, below = of tc] {
    Reports identified from databases (n=1090):\\
    \>\>\>\>\>ACM (n=81)\\
    \>\>\>\>\>IEEE (n=194)\\
   \>\>\>\>\>Google Scholar (n=139)\\
   \>\>\>\>\>Proquest (n=250)\\
    \>\>\>\>\>Scopus (n=298)\\
    \>\>\>\>\>WebOfScience (n=128)\\
};

\node (n3)  [mynode, join]   { Reports screened based on title and abstracts (n=1090) };
\node (n4)  [mynode, join]   { Reports full-text retrieved (n=422) };
\node (n5)  [mynode, join]   { Reports assessed for eligibility (n=414) };
\node (n6)  [mynode, join]   { Studies included (n=132) };

\node (n3r) [mynode, right=of n3]    {
    Reports excluded (n=668):\\
    \>\>\>\>\>Automated duplication detected (n=290)\\
    \>\>\>\>\>Not available in English (n=6)\\
   \>\>\>\>\>Not matched eligibility themes  (n=372)\\
};
    
\node (n4r) [mynode, right=of n4]    { 
    Reports excluded (n=8):\\
    \>\>\>\>\>Manual duplication detected (n=6)\\
    \>\>\>\>\>Full-text not available (n=2)\\
 };
 
\node (n5r) [mynode, right=of n5]    {
    Reports excluded (n=282):\\
    \>\>\>\>\>Not matched eligibility themes (n=157)\\
    \>\>\>\>\>Missing research method (n=113)\\
    \>\>\>\>\>Insufficient report quality (n=12)\\
};

\draw[arrow] (n3) -- (n3r);
\draw[arrow] (n4) -- (n4r);
\draw[arrow] (n5) -- (n5r);

\begin{scope}[node distance=7mm]
    \node(l1)[mylabel, minimum height=3.3cm, below left=-2mm and 8mm of n2.north west] {\rotatebox{90}{Identification}};
    \node(l2)[mylabel, minimum height=6.5cm]  {\rotatebox{90}{Screening}};
    \node(l3)[mylabel, minimum height=0.8cm]  {\rotatebox{90}{Included}};
\end{scope}

\end{tikzpicture}


 }
\caption{\label{fig:prisma} PRISMA chart to show the number of articles included/excluded at each stage of the review.}
\figalttext{PRISMA flow diagram for the review, with three stages on the left (Identification, Screening, Included) and exclusion counts on the right. From 1,090 records identified across five databases and Google Scholar, 290 duplicates and 6 non-English reports were removed, leaving 1,090 reports screened by title and abstract. After excluding 372 ineligible records and 2 with no full text, 422 reports were retrieved and 414 assessed for eligibility; 289 were excluded at full-text screening, leaving 132 studies included.}
\end{figure}

\subsubsection{Demographics}
\label{sec:demog}
This section details the demographics of the 132 included studies, which are all catalogued with their methodological details in Appendix~\ref{appdx:scope}. Table~\ref{tab:demo_year} shows the distribution of the frequency of publications for each year, showing two significant waves of publications in 2020 and 2023. This may be a coincidence or a result of the restrictions imposed during the \mbox{COVID-19} pandemic, but this was not evident from the literature.
In terms of the types of publication (Table~\ref{tab:demo_type}), the majority were journal articles (n=78), followed by conference papers (n=34) and PhD theses (n=19), with two book and one technical report. Even though we included grey literature in our searches, only three instances met our inclusion criteria. That is because many reports (grey literature) did not describe the employed research methodology, or their descriptions lacked basic details and could be contested by a reviewer.

To describe the types of studies reviewed, summary statistics are provided about the research methods used (e.g. case study, qualitative, quantitative, literature review etc), data collection methods and sample sizes. Starting with research methods (Table~\ref{tab:demo_resmethod}), we encountered an overlapping mixture of designs: 71\% of studies (n=94) employed qualitative methods, 42\% (n=56) used quantitative methods, and 27\% (n=36) reported findings from case study analyses. Of the studies that used quantitative methods, only six conducted a thorough statistical analysis; the others were limited to reporting the frequencies associated with their variables of interest.
As for data collection methods, we observed a combination of surveys, interviews and literature review. Despite 27 studies citing existing datasets to introduce their research, all studies focused on primary analysis (original data collection), with no studies conducting a secondary analysis of existing data. Apropos sample sizes, Table~\ref{tab:demo_samplesize} presents a grouped view of the sample sizes from those studies that reported a numeric sample. Just over 40\% (40\%) of these studies (n=38) had a sample size of $<25$.
It is worth noting that the literature presented a substantial focus on interventions intended to reduce cybersecurity risks. Overall, 64 (48\%) of the studies reviewed proposed frameworks, of which 28 employed surveys to validate the applicability and efficiency of the proposed intervention.

Lastly, we also coded the geographical regions examined in the literature. Despite many reports claiming that developing countries were underrepresented, we encountered a substantial body of research conducted in them. Of the 114 studies explicitly stating the study location, 37 (32\%) investigated developing countries. They concentrated on the Middle East, South East Asia and some African countries, leaving areas such as South America largely uncovered. The discussion section (section~\ref{sec:developing} and Fig.~\ref{fig:geodistrib}) presents a more detailed analysis of the geographical coverage and its challenges.

\begin{table}[!htbp]
\begin{minipage}[t]{0.48\textwidth}
\centering
\small
\caption{Distribution of studies by publication year.}
\label{tab:demo_year}
\begin{tabular}{ r | l }
            \hline
            Pub. year & Freq. \\
            \hline
                2017 & 7 \\
                2018 & 13 \\
                2019 & 18 \\
                2020 & 33 \\
                2021 & 19 \\
                2022 & 9 \\
                2023 & 27 \\
                2024 & 6 \\
           \hline
         \end{tabular}
\end{minipage}
\hfill
\begin{minipage}[t]{0.48\textwidth}
\centering
\small
\caption{Distribution of studies publication type.}
\label{tab:demo_type}
\begin{tabular}{r |l}
                \hline
                Publication type & Freq. \\
                \hline
                Journal article & 78 \\
                Conference paper & 34 \\
                PhD thesis & 19 \\
                Book & 2 \\
                Technical report & 1 \\
             \hline
            \end{tabular}
\end{minipage}
\par\medskip
\begin{minipage}[t]{0.48\textwidth}
\centering
\small
\caption{Distribution of research methods.}
\label{tab:demo_resmethod}
\begin{tabular}{ r | l }
             \hline
                Research method & Freq. \\
             \hline
            Qualitative & 94 \\
           Quantitative & 56 \\
            Case Study & 36 \\
            Literature Review & 1 \\
             \hline
            \end{tabular}
\end{minipage}
\hfill
\begin{minipage}[t]{0.48\textwidth}
\centering
\small
\caption{Distribution of sample sizes on primary (surveys) analysis.}
\label{tab:demo_samplesize}
\begin{tabular}{ r | l }
             \hline
                Sample size & Freq. \\
             \hline
            1 to 10 & 21 \\
            11 to 25 & 21 \\
            26 to 50 & 12 \\
            51 to 100 & 8 \\
            101 to 200 & 13 \\
            201 to 300 & 6 \\
            301 to 400 & 4 \\
            401 to 500 & 3 \\
            501 to 600 & 3 \\
            601 to 800 & 2 \\
            801 to 5000 & 1 \\
             \hline
            \end{tabular}
\end{minipage}
\end{table}

\section{Results}
\label{sec:resuls}
The data extraction and thematic analysis resulted in the identification of 44 themes across 2989 quotations.  Space limitations preclude a discussion of all 44 topics.  Consequently, we concentrate on those themes that were discussed in at least ten studies, either to motivate the study (i.e. drivers) or to report findings about that theme (excluded themes were discussed in an average of 3 studies, range=1--9).  Examples of themes not addressed in this paper include: discussions of how existing legislation and standards do not account for SME needs; that SMEs typically do not report cyberattacks suffered; that COVID-19 forced digitalisation amongst SMEs; and that risk perception increases motivation towards cybersecurity.
Table~\ref{tab:themes-tb} lists the themes and the number of studies that discussed them as drivers (to motivate the study) or conclusions (where new findings were reported), along with citations to the studies that discussed them (see section~\ref{sec:gap-analysis}).
Cells in the table are shaded to reflect the number of studies discussing each theme. The higher the frequency, the darker the colour. The last column ($\cap$) represents the number of studies for which that theme was both a driver and a conclusion. These studies are perhaps particularly interesting as they discuss the theme when introducing the research and subsequently test its associated predictions.
An extended version of Table~\ref{tab:themes-tb} that includes the complete set of themes (i.e. those not discussed here) is available at~\url{https://osf.io/ps7xy/}.

\begin{table*}[!htbp]
\caption{Synthesised themes that emerged in more than 10 studies.\label{tab:themes-tb}}
 \small	
\begin{tabular}
{
    | >{\raggedright\arraybackslash} m{.14\textwidth} 
    | >{\centering} m{.04\linewidth} | m{.28\linewidth} 
    | >{\centering} m{.04\linewidth} | m{.28\linewidth} 
    | >{\centering} m{.04\linewidth} 
    |  c |
}

\hline
    \textbf{Synthesised} &  
    \multicolumn{2}{c|}{\textbf{Drivers}} & 
    \multicolumn{2}{c|}{\textbf{Conclusion}} & 
    \multirow{2}{*}{\textbf{Total}} &
     \multirow{2}{*}{\textbf{ $\cap$ }} \\
    \textbf{Theme}& \centering{\textbf{No.}} & \centering{\textbf{Studies}} & \centering{\textbf{No.}} & \centering{\textbf{Studies}} & & \\

\hline

Lack of awareness & \cellcolor{black!41}41 & \cite{White2020} \cite{Lejaka2019} \cite{Dykstra2020} \cite{Ozkan2019} \cite{Loffler2021} \cite{Elezaj2019} \cite{Rae2019} \cite{Mitrofan2020} \cite{Rae2020} \cite{Bada2019} \cite{Nasir2020} \cite{Bailetti2020} \cite{Ahmed2021} \cite{Kabanda2018} \cite{Ikuero2022} \cite{Pickering2021} \cite{Johannsen2020} \cite{Almubayedh2018} \cite{Carias2020} \cite{Alharbi2021} \cite{Gafni2019} \cite{Yudhiyati2021} \cite{Saban2021} \cite{Gamisch2023ASecurity} \cite{Chidukwani2022} \cite{Ogunyebi2018AnSector} \cite{Mmango2023CyberICECET} \cite{Asprion2023} \cite{Diaz-Piraquive2023CybersecurityColombia} \cite{Shojaifar2023DesignTool} \cite{Harting2023} \cite{Almoaigel2023ImplementationSMEs} \cite{Moeti2023InformationEnterprise} \cite{Eybers2022InvestigatingSMEs} \cite{Neri2024} \cite{Al-Somali2024OrganizationalCulture} \cite{Shojaifar2020SMEsSharing} \cite{Oroni2023StructuralPerformance} \cite{Ponsard2019SurveySMEs} \cite{Scholl2023SustainableEffect} \cite{Heidt2018TheSecurity} & \cellcolor{black!30}30 & \cite{Huaman2021} \cite{Ncubukezi2020} \cite{Mokwetli2018} \cite{White2020} \cite{Heidenreich2019} \cite{Alahmari2020} \cite{Rae2019} \cite{Rae2020} \cite{Bada2019} \cite{Teymourlouei2019} \cite{Ahmed2021} \cite{Kabanda2018} \cite{Kljucnikov2019} \cite{Alahmari2021} \cite{Heidt2019} \cite{Pickering2021} \cite{Osborn2018} \cite{Ulrich2020} \cite{Yudhiyati2021} \cite{Erdogan2023} \cite{Rawindaran2023EnhancingKingdom} \cite{Ncubukezit2022} \cite{Almoaigel2023ImplementationSMEs} \cite{Nkhoma2018InformationSurveys} \cite{Eybers2022InvestigatingSMEs} \cite{Neri2024} \cite{Sia2021ReasonsStudies} \cite{Oroni2023StructuralPerformance} \cite{Scholl2023SustainableEffect} \cite{Heidt2018TheSecurity} & 57 & 14 \\ 
\hline

Under-funded and resource-constrained cybersecurity programs & \cellcolor{black!42}42 & \cite{White2020} \cite{Barlette2017} \cite{Heidenreich2019} \cite{Mierzwa2017} \cite{Ozkan2019} \cite{Loffler2021} \cite{Elezaj2019} \cite{Rae2019} \cite{Mitrofan2020} \cite{Rae2020} \cite{Bada2019} \cite{Ahmed2021} \cite{Barlette2019} \cite{Heidt2019} \cite{Chandra2020} \cite{Pickering2021} \cite{Ozkan2020} \cite{Carias2020} \cite{Young2020} \cite{Bountouni2023AStudy} \cite{Auyporn2023AStandard} \cite{Chidukwani2022} \cite{Berry2018AnThreats} \cite{Cartwright2023CascadingCompanies} \cite{Mmango2023CyberICECET} \cite{Tsiodra2023CyberStudy} \cite{Erdogan2023} \cite{Asprion2023} \cite{Franco2023CyberTEAInvestment} \cite{Shojaifar2023DesignTool} \cite{Rawindaran2023EnhancingKingdom} \cite{Roy2023FrameworkInCACCT} \cite{Jamil2024Human-centricBehaviours} \cite{Pickering2023ITechnology} \cite{Almoaigel2023ImplementationSMEs} \cite{Eybers2022InvestigatingSMEs} \cite{Wilson2023} \cite{Neri2024} \cite{Al-Somali2024OrganizationalCulture} \cite{Aigbefo2022TheIntention} \cite{Heidt2018TheSecurity} \cite{Lejaka2023TowardsSMMEs} & \cellcolor{black!19}19 & \cite{Heidenreich2019} \cite{Dykstra2020} \cite{Alahmari2020} \cite{Kabanda2018} \cite{Kljucnikov2019} \cite{Alahmari2021} \cite{Heidt2019} \cite{Johannsen2020} \cite{VanHaastrecht2021} \cite{Osborn2018} \cite{Ulrich2020} \cite{Tam2021} \cite{Chidukwani2022} \cite{Cartwright2023CascadingCompanies} \cite{Almoaigel2023ImplementationSMEs} \cite{Nkhoma2018InformationSurveys} \cite{Eybers2022InvestigatingSMEs} \cite{Neri2024} \cite{Heidt2018TheSecurity} & 53 & 8 \\ 
\hline

Limited literacy & \cellcolor{black!39}39 & \cite{White2020} \cite{Barlette2017} \cite{Heidenreich2019} \cite{Alahmari2020} \cite{Loffler2021} \cite{Rae2019} \cite{Rae2020} \cite{Bada2019} \cite{Raineri2020} \cite{Ahmed2021} \cite{Barlette2019} \cite{Brunner2018} \cite{Chandra2020} \cite{Johannsen2020} \cite{McLilly2020} \cite{Almubayedh2018} \cite{Young2020} \cite{Alharbi2021} \cite{Gafni2019} \cite{Dimitrov2019} \cite{Bountouni2023AStudy} \cite{Ogunyebi2018AnSector} \cite{Berry2018AnThreats} \cite{Cartwright2023CascadingCompanies} \cite{Carias2021CyberSMEs} \cite{Tsiodra2023CyberStudy} \cite{Asprion2023} \cite{Diaz-Piraquive2023CybersecurityColombia} \cite{Shojaifar2023DesignTool} \cite{Rawindaran2023EnhancingKingdom} \cite{Roy2023FrameworkInCACCT} \cite{Ncubukezit2022} \cite{Jamil2024Human-centricBehaviours} \cite{Pickering2023ITechnology} \cite{Almoaigel2023ImplementationSMEs} \cite{Ponsard2020MethodologyBelgium} \cite{Neri2024} \cite{Shojaifar2020SMEsSharing} \cite{Lejaka2023TowardsSMMEs} & \cellcolor{black!19}19 & \cite{Mokwetli2018} \cite{White2020} \cite{Dykstra2020} \cite{Mitrofan2020} \cite{Alahmari2021} \cite{Heidt2019} \cite{Pickering2021} \cite{Osborn2018} \cite{Ulrich2020} \cite{Tam2021} \cite{Yudhiyati2021} \cite{Chidukwani2022} \cite{Cartwright2023CascadingCompanies} \cite{Asprion2023} \cite{Pickering2023ITechnology} \cite{Almoaigel2023ImplementationSMEs} \cite{Ruhwanya2019InformationTanzania} \cite{Neri2024} \cite{Heidt2018TheSecurity} & 52 & 6 \\ 
\hline

Rising cyberattacks against SME & \cellcolor{black!41}41 & \cite{White2020} \cite{Barlette2017} \cite{Carias2021} \cite{Dykstra2020} \cite{Alahmari2020} \cite{Ozkan2019} \cite{Loffler2021} \cite{Mitrofan2020} \cite{Ahmed2021} \cite{Ikuero2022} \cite{Barlette2019} \cite{Heidt2019} \cite{Johannsen2020} \cite{McLilly2020} \cite{VanHaastrecht2021} \cite{Tam2021} \cite{Alharbi2021} \cite{Dimitrov2019} \cite{Yudhiyati2021} \cite{Saban2021} \cite{Gamisch2023ASecurity} \cite{Auyporn2023AStandard} \cite{Chidukwani2022} \cite{Ogunyebi2018AnSector} \cite{Rodriguez-Baca2022BusinessSMEs} \cite{Cartwright2023CascadingCompanies} \cite{Mmango2023CyberICECET} \cite{Tsiodra2023CyberStudy} \cite{Erdogan2023} \cite{Asprion2023} \cite{Harting2023} \cite{Jamil2024Human-centricBehaviours} \cite{Almoaigel2023ImplementationSMEs} \cite{Azinheira2023} \cite{Ponsard2020MethodologyBelgium} \cite{Neri2024} \cite{Al-Somali2024OrganizationalCulture} \cite{Cruzado2022Reference27032} \cite{Ponsard2019SurveySMEs} \cite{Scholl2023SustainableEffect} \cite{Lejaka2023TowardsSMMEs} & \cellcolor{black!1}0 & - & 41 & 0 \\ 
\hline

Lack of tailored solutions and frameworks & \cellcolor{black!28}28 & \cite{Park2021} \cite{Barlette2017} \cite{Heidenreich2019} \cite{Carias2021} \cite{Mierzwa2017} \cite{Ozkan2019} \cite{Rae2019} \cite{Rae2020} \cite{Nasir2020} \cite{Ahmed2021} \cite{Alghamdi2019} \cite{Chandra2020} \cite{Pickering2021} \cite{Johannsen2020} \cite{Ozkan2020} \cite{Katt2018} \cite{McLilly2020} \cite{VanHaastrecht2021} \cite{Osborn2018} \cite{Carias2020} \cite{Bountouni2023AStudy} \cite{Carias2021CyberSMEs} \cite{Erdogan2023} \cite{Asprion2023} \cite{Jamil2024Human-centricBehaviours} \cite{Azinheira2023} \cite{Ruhwanya2019InformationTanzania} \cite{AL-Dosari2023Risk-ManagementApproach} & \cellcolor{black!14}14 & \cite{Park2021} \cite{Heidenreich2019} \cite{Lejaka2019} \cite{Ozkan2019} \cite{Rae2019} \cite{Kabanda2018} \cite{VanHaastrecht2021} \cite{Tam2021} \cite{Yudhiyati2021} \cite{Asprion2023} \cite{Pickering2023ITechnology} \cite{Azinheira2023} \cite{Ruhwanya2019InformationTanzania} \cite{Ponsard2019SurveySMEs} & 34 & 8 \\ 
\hline

Literature overlooked SMEs specific needs & \cellcolor{black!25}25 & \cite{Barlette2017} \cite{Heidenreich2019} \cite{Carias2021} \cite{Dykstra2020} \cite{Alahmari2020} \cite{Ozkan2019} \cite{Rae2019} \cite{Mitrofan2020} \cite{Rae2020} \cite{Nasir2020} \cite{Ahmed2021} \cite{Ikuero2022} \cite{Alahmari2021} \cite{Heidt2019} \cite{Durowoju2020} \cite{Perez-Gonzalez2019} \cite{VanHaastrecht2021} \cite{Osborn2018} \cite{Jantti2020} \cite{Alharbi2021} \cite{Yudhiyati2021} \cite{Auyporn2023AStandard} \cite{Chidukwani2022} \cite{Berry2018AnThreats} \cite{Wilson2023} & \cellcolor{black!9}9 & \cite{Park2021} \cite{Alahmari2020} \cite{Bada2019} \cite{Kabanda2018} \cite{Brunner2018} \cite{Alahmari2021} \cite{Tam2021} \cite{Auyporn2023AStandard} \cite{Almoaigel2023ImplementationSMEs} & 31 & 3 \\ 
\hline

Low risk perception & \cellcolor{black!22}22 & \cite{Barlette2017} \cite{Mierzwa2017} \cite{Elezaj2019} \cite{Rae2019} \cite{Rae2020} \cite{Ahmed2021} \cite{Kabanda2018} \cite{Ikuero2022} \cite{Alahmari2021} \cite{Jantti2020} \cite{Young2020} \cite{Gafni2019} \cite{Yudhiyati2021} \cite{Saban2021} \cite{Ogunyebi2018AnSector} \cite{Asprion2023} \cite{Franco2023CyberTEAInvestment} \cite{Jamil2024Human-centricBehaviours} \cite{Eybers2022InvestigatingSMEs} \cite{Wilson2023} \cite{Ponsard2020MethodologyBelgium} \cite{Ponsard2019SurveySMEs} & \cellcolor{black!18}18 & \cite{Dykstra2020} \cite{Saa2017} \cite{Rae2019} \cite{Mitrofan2020} \cite{Rae2020} \cite{Ahmed2021} \cite{Kabanda2018} \cite{Kljucnikov2019} \cite{Alahmari2021} \cite{Heidt2019} \cite{Osborn2018} \cite{Almubayedh2018} \cite{Ulrich2020} \cite{Yudhiyati2021} \cite{Saban2021} \cite{Chidukwani2022} \cite{Jamil2024Human-centricBehaviours} \cite{Wilson2023} & 31 & 9 \\ 
\hline

Overwhelmed or sparse or non-existent cybersecurity leadership & \cellcolor{black!13}13 & \cite{Alahmari2020} \cite{Loffler2021} \cite{Mitrofan2020} \cite{Bada2019} \cite{Barlette2019} \cite{Chandra2020} \cite{Mayer2017} \cite{McLilly2020} \cite{Ahmed2020} \cite{Gafni2019} \cite{Almoaigel2023ImplementationSMEs} \cite{Marett2019InformationAnalysis} \cite{Ponsard2019SurveySMEs} & \cellcolor{black!16}16 & \cite{White2020} \cite{Dykstra2020} \cite{Mitrofan2020} \cite{Barlette2019} \cite{Kljucnikov2019} \cite{Alahmari2021} \cite{Heidt2019} \cite{Pickering2021} \cite{Mayer2017} \cite{Osborn2018} \cite{Carias2020} \cite{Tam2021} \cite{Saban2021} \cite{Erdogan2023} \cite{Wilson2023} \cite{Heidt2018TheSecurity} & 26 & 3 \\ 
\hline

Poor security operations & \cellcolor{black!10}10 & \cite{Heidenreich2019} \cite{Rawindaran2021} \cite{Alahmari2021} \cite{Jantti2020} \cite{Ulrich2020} \cite{Yudhiyati2021} \cite{Bountouni2023AStudy} \cite{Harting2023} \cite{Eybers2022InvestigatingSMEs} \cite{Neri2024} & \cellcolor{black!11}11 & \cite{Ncubukezi2020} \cite{Dykstra2020} \cite{Rae2019} \cite{Ikuero2022} \cite{Douchek2020} \cite{Carias2020} \cite{Tam2021} \cite{Yudhiyati2021} \cite{Chidukwani2022} \cite{Diaz-Piraquive2023CybersecurityColombia} \cite{Neri2024} & 19 & 2 \\ 
\hline

Increasing financial loss due to cyber-attacks & \cellcolor{black!17}17 & \cite{Park2021} \cite{Huaman2021} \cite{Dykstra2020} \cite{Alahmari2020} \cite{Elezaj2019} \cite{Mitrofan2020} \cite{Teymourlouei2019} \cite{Raineri2020} \cite{Ahmed2021} \cite{Alahmari2021} \cite{Heidt2019} \cite{Durowoju2020} \cite{McLilly2020} \cite{Almubayedh2018} \cite{Ulrich2020} \cite{Mmango2023CyberICECET} \cite{Aigbefo2022TheIntention} & \cellcolor{black!2}2 & \cite{Alahmari2020} \cite{Rae2020} & 18 & 1 \\ 
\hline

Cloud adoption and managed services minimise cybersecurity challenges & \cellcolor{black!9}9 & \cite{Dykstra2020} \cite{Feher2020} \cite{Saa2017} \cite{Mitrofan2020} \cite{Johannsen2020} \cite{Ozkan2020} \cite{McLilly2020} \cite{Jantti2020} \cite{Carias2020} & \cellcolor{black!6}6 & \cite{White2020} \cite{Alahmari2020} \cite{Feher2020} \cite{Ahmed2021} \cite{Alahmari2021} \cite{Ahmed2020} & 14 & 1 \\ 
\hline


Legislation and informative content improve cybersecurity & \cellcolor{black!5}5 & \cite{Elezaj2019} \cite{Bada2019} \cite{Ikuero2022} \cite{Gafni2019} \cite{Saban2021} & \cellcolor{black!7}7 & \cite{Huaman2021} \cite{Carias2021} \cite{Alahmari2020} \cite{Raineri2020} \cite{Saban2021} \cite{Rawindaran2023EnhancingKingdom} \cite{Ponsard2020MethodologyBelgium} & 11 & 1 \\ 
\hline

\end{tabular}
\par\smallskip
\begin{minipage}{\textwidth}
\footnotesize Column colours represent the frequency of occurrence; the darker the colour, the higher the value.
\end{minipage}
\end{table*}

The following subsections discuss each theme, starting with an overview of the topic.

\subsection{Lack of awareness of cybersecurity risks}
\label{res:awareness}

The \emph{lack of awareness of cybersecurity risks} emerged as the most frequent theme in the literature. It was discussed in 41 studies as a motivation (driver) for the research, while 30 studies presented novel findings (conclusions). This profile indicates that recent empirical studies continue to suggest that SMEs lack awareness of cybersecurity threats.
While there is symmetry in the volume of driver and conclusion themes, they emerged from different rationales. For instance, decision-makers lacking awareness emerged as a driver in nine studies~\cite{Loffler2021, Saban2021, Bailetti2020, Ahmed2021, Yudhiyati2021,  Ogunyebi2018AnSector, Asprion2023, Scholl2023SustainableEffect, Aigbefo2022TheIntention}, but was a conclusion in only four~\cite{Kljucnikov2019, Alahmari2021, Oroni2023StructuralPerformance, Heidt2018TheSecurity}.
Conversely, lack of awareness across all employees emerged in the introduction section of three studies~\cite{Loffler2021, Elezaj2019, Johannsen2020} and from original findings presented in four~\cite{Huaman2021, Bada2019, Ulrich2020, Ncubukezit2022}. 

Several empirical studies concluded that employees were aware of cybersecurity dangers, but because their leadership was not, cybersecurity initiatives were not prioritised or were ignored~\cite{Saban2021, Loffler2021, Ahmed2021, Kljucnikov2019, Alahmari2021, Yudhiyati2021}. Conversely, five empirical studies~\cite{Huaman2021, Loffler2021, Elezaj2019, Johannsen2020, Ulrich2020} concluded that only leadership and tech departments had awareness and that this was not propagated to the remainder of the organisation. One of these studies suggested that this situation made SMEs more vulnerable to phishing attacks~\cite{Huaman2021}.

A \emph{lack of awareness of existing solutions} has also gained attention in the literature. It emerged as a motivation in two studies~\cite{Ozkan2019, Diaz-Piraquive2023CybersecurityColombia} and as a conclusion in five others~\cite{Alahmari2020, Rae2019, Rae2020, Teymourlouei2019, Heidt2019}. Each of these studies suggested that SMEs could not establish effective cybersecurity defences because they did not employ available solutions, and that this was because they were unaware of them.

A \emph{lack of awareness of digital assets} (e.g. infrastructure, customer data or digital monies) surfaced only as a finding (i.e. it was not discussed in the introduction sections of any studies reviewed), suggesting that it is a recent discovery. Three studies reported that SMEs were aware of such threats. Still, because they underestimated their asset value, they perceived themselves as unlikely cyberattack targets and considered their primary defences `good enough'~\cite{White2020, Osborn2018, Ulrich2020}.  
A \emph{lack of awareness of regulatory cybersecurity requirements} (i.e. compliance) emerged as a driver in two studies~\cite{Rae2019, Nasir2020}. However, no research was identified that tested this. In addition to digital attack vectors, a \emph{lack of awareness of non-technical threats} emerged as a conclusion in (only) one study~\cite{Mokwetli2018}. According to this research, in the context of SMEs, humans have become one of the main vectors for data breaches because attention has been too focused on technological controls against cyber threats at the cost of disregarding human errors.

Various studies that surveyed SMEs revealed that increasing cybersecurity awareness amongst employees increases the likelihood of the (self-reported) adoption of appropriate security behaviours, thereby improving the organisation's cybersecurity posture~\cite{Saban2021, Mokwetli2018, Alahmari2020, Raineri2020, Kljucnikov2019, Heidt2019, Pickering2021, VanHaastrecht2021}. A SR by \citeauthor{VanHaastrecht2021} summarises this nicely: `\emph{Furthering human knowledge and improving the technical cybersecurity posture of an SME go hand-in-hand}'~\cite{VanHaastrecht2021}.

It is worth noting that multiple empirical studies suggested that SMEs are not fully aware of the cybersecurity risks they face~\cite{Ncubukezi2020, Heidenreich2019, Ahmed2021, Kabanda2018, Pickering2021, Yudhiyati2021, Kljucnikov2019, Alahmari2021, Huaman2021, Bada2019, Ulrich2020, Osborn2018, Oroni2023StructuralPerformance, Moeti2023InformationEnterprise}, or the solutions available to them~\cite{Alahmari2020, Rae2019, Rae2020, Teymourlouei2019, Heidt2019}. However, many of these studies failed to enumerate and specify the risks, threats, or solutions they referred to. Instead, they often grouped various classes of risks (e.g., internal threats, ransomware, phishing attacks) under the umbrella term `cybersecurity risks'. One study even suggested that SMEs face the same threats as large companies, but without enumerating what these specific threats were~\cite{Al-Somali2024OrganizationalCulture}.

\subsection{Under-funded and resource-constrained cybersecurity programmes}
\label{res:underfunded}

A total of 53 studies discussed under-investment and resource constraints as barriers to effective cybersecurity in SMEs. Of these, 43 papers touched on this subject to introduce their research, while 19 papers drew conclusions (from an empirical assessment) supporting this theme.  In both cases, however, discussions were often generic and lacked important detail compared to other themes, a situation that is further discussed in section~\ref{disc:lackprogress}.

There was a clear consensus across the literature that under-investment in cybersecurity is an endemic pitfall for SMEs. Still, there was no agreement on the causes or extent of the problem. While studies claim SMEs do not invest enough in cybersecurity, none provided recommendations as to what sufficient levels of investment would be. This situation was also observed by \citeauthor{Ahmed2021}, who emphasised the need for guidance to inform what adequate levels of financial allocation for cybersecurity are~\cite{Ahmed2021}.  Furthermore, only two studies~\cite{Heidt2019, Osborn2018} investigated how much SMEs invested in cybersecurity before concluding that it was insufficient. One of the few studies that actually reported investments analysed findings from a survey of UK-based SMEs.  Their results suggested that SMEs invested less than £1000 a month on cybersecurity, despite recognising that they were unprepared to defend against cyber threats~\cite{Heidt2019}.
Similar results emerged from another study of UK SMEs, which found (through interviews) that cybersecurity investments did not scale with company size and were capped at £10,000  per year~\cite{Osborn2018}. 
In contrast, in a study of SMEs in Tanzania, the authors reported that SMEs invested proportionally more in cybersecurity than do larger organisations~\cite{Ruhwanya2019InformationTanzania}. However, no actual investment figures were reported in that study to verify their conclusion.

Some studies suggest that SMEs have financial resources but fail to allocate them to cybersecurity programmes. A reason commonly stated for this was that decision-makers lacked awareness of the cyber threats they faced (as discussed in section~\ref{res:awareness}) or leaders did not have enough literacy to understand how to invest in cybersecurity (e.g.~\cite{Chidukwani2022}). This explanation was discussed in the introduction of four studies~\cite{Mitrofan2020, Barlette2019, Orellana2020, Franco2023CyberTEAInvestment} and was supported by the findings of three other studies that interviewed SMEs based in Germany~\cite{Heidt2018TheSecurity}, Italy~\cite{Neri2024} and Western Europe~\cite{Heidt2019}: all developed countries. A review of the literature pertaining to the cybersecurity awareness of South African SMEs drew the opposite conclusion, suggesting that the needs of SMEs in developing countries are unique and that they have limited access to financial resources~\cite{Lejaka2023TowardsSMMEs}.  Another study of SMEs in South Africa attributed the under-investment problem to the intangible benefits and unclear returns on investment in cybersecurity, drawing a parallel between illiterate decision-makers and the specific challenges faced by SMEs in developing countries~\cite{Kabanda2018}.

There was also a discussion in the literature that governments and private companies provide subsidies (dedicated to SMEs) for cybersecurity development. However, these often remain unused due to SMEs not being aware of the support available to them~\cite{Heidt2019, Heidt2018TheSecurity}.

Despite various studies discussing SME resource constraints, only a few specified the consequences of such limitations. One paper postulated that SMEs could not access solutions and external support due to a shortage of funds~\cite{Barlette2019}. Another supported (empirical findings) this statement, suggesting that commercial solutions are too expensive for SMEs~\cite{Bountouni2023AStudy}. Along the same lines, the authors of another study reported that Thailand-based SMEs struggled to adopt cybersecurity standards due to limited financial resources. Even though investigated standards (e.g. ISO 27001, ENISA, NIST CSF, UK Cyber Essentials) were accessible (less than £250) or free of charge, their adoption required resources beyond their reach~\cite{Auyporn2023AStandard}. This issue remains valid, even for standards and regulations that SMEs are obliged to comply with, such as the Payment Card Industry Data Security Standard (PCI-DSS) and the General Data Protection Regulation (GDPR)~\cite{Chidukwani2022}.
Two studies claimed (conclusion) that financial constraints resulted from their decision-makers being unaware of cybersecurity issues/needs and consequently under-investing in security measures~\cite{Eybers2022InvestigatingSMEs}.

Contrary to perhaps popular opinion, no research has observed that cybersecurity budgets substantially increase after suffering a cyberattack~\cite{Rae2020}.

\subsection{Limited cybersecurity literacy}
\label{res:literacy}

\emph{Limited cybersecurity literacy} concerns an individual's competence and domain knowledge in cybersecurity, whereas  \emph{lack of cybersecurity awareness} as previously discussed in section~\ref{res:awareness} is about not knowing of the existence of cyber-risks and -threats.

\emph{Limited cybersecurity literacy} emerged as a recurring theme in the literature; it surfaced as a motivation in one third of the articles reviewed (n=39) and was corroborated in 19 studies. Despite being frequently discussed, most studies mentioned it generically, failing to pinpoint which cybersecurity domain knowledge was missing and to specify which roles lacked literacy. Consequently, only a few details emerged on this subject, highlighting the need for more comprehensive and targeted research to investigate shortfalls in this type of literacy.

Research motivated by this theme referred to previous discoveries suggesting that SME leaders and decision-makers presented limited literacy on cybersecurity~\cite{Barlette2017, Ahmed2021, Barlette2019, Ogunyebi2018AnSector}. This notion was tested and confirmed by two other studies~\cite {Ruhwanya2019InformationTanzania, Heidt2018TheSecurity}.
One study referenced previous research to suggest that insufficient cybersecurity proficiency among SME leadership has led to inefficient strategies and scarce investments in cybersecurity initiatives~\cite{Ahmed2021}. 
Studies indicated that SMEs' limited cybersecurity literacy resulted from unprepared cybersecurity professionals who failed to engage with other employees. This lack of preparedness was reported to have resulted from a combination of a lack of higher education and available training curricula that did not match industry needs. This emerged from novel findings of one study~\cite{Ulrich2020} , and as a motivation for the other two studies~\cite{Dimitrov2019, Lejaka2019}.

Research using interviews to investigate levels of cybersecurity awareness across SMEs in the North American medical industry reported that SME personnel literacy was directly related to the organisation's financial capacity; the more prosperous the SME, the higher its overall literacy level~\cite{Dykstra2020}.

In two studies, the authors presented data to conclude that limited literacy and resource constraints in relation to cybersecurity (discussed in more detail below) form a vicious circle, where decision-makers with little to no literacy in cybersecurity invest less in cybersecurity, consequently limiting their access to skilled talent~\cite{Chidukwani2022, Heidt2018TheSecurity}. Along the same lines, one study concluded that a lack of cybersecurity awareness training substantially hindered SMEs' cyber resilience, and that this was caused by budgetary constraints~\cite{Eybers2022InvestigatingSMEs}.

\subsection{Lack of tailored solutions and frameworks}
\label{res:solutions}

One quarter of the literature touched on this theme, and studies uniformly agreed that existing solutions were unsuitable for SMEs.  However, only three papers attempted to describe the available solutions, and no articles described why the requirements of SMEs were unmet by existing solutions.
Moreover, we did not encounter any mention of an existing taxonomy of solutions, nor was there any research attempting to build one. Instead, authors tended to discuss more `cybersecurity solutions', referring to different classes of solutions such as standards, regulations, frameworks, software, and technologies.

As mentioned, only three studies described the existing solutions as part of their research.  In each case, the authors reported theoretical investigations of information security risk management and cybersecurity governance standards such as ISO 27001, the NIST Cyber Security Framework (CSF) and the UK's Cyber Essentials scheme~\cite{Azinheira2023, Asprion2023, Ponsard2019SurveySMEs}. They concluded that they were unsuitable for SMEs. 
\citeauthor{Ponsard2019SurveySMEs}'s conclusion summarises well other studies' findings and mentions where this theme emerged from:
`\emph{Unfortunately, standards and frameworks were designed in the first place with large companies in mind and are not tailored to the needs and resources of SMEs. By definition, SMEs are also very heterogeneous, and a single solution cannot fit them all. Consequently, the need for a comprehensive, flexible, and cost-minded framework is clear, and major actors such as the European Union and other major standards are beginning to work on it}'~\cite{Ponsard2019SurveySMEs}.

The studies that motivated their research on this issue (but did not report novel findings) agreed on the challenges and reasons for a lack of tailored solutions and frameworks. For example, \citeauthor{Alghamdi2019} quoted previous research that attributed the problem to the resources needed to implement existing solutions being beyond the reach of SMEs~\cite{Alghamdi2019}. 
Four studies presented more details on their motivation, stating that existing solutions were overly complicated for SMEs' constrained resources (human, financial, and silicon)~\cite{Barlette2017, Carias2021, Alghamdi2019, Katt2018}. At the same time, three studies collected data and reported findings that supported this conclusion~\cite{Rae2019, Ruhwanya2019InformationTanzania, Bountouni2023AStudy}. According to all of these studies, solutions were designed to address the needs of larger organisations, and given the different requirements and capabilities of SMEs, solutions were considered unfit. \citeauthor{Ruhwanya2019InformationTanzania} findings suggested that the implementation of existing frameworks required a certain cybersecurity maturity --- such as the existence of a well-defined information security policy and the presence of information security officers --- that exceeds the capabilities of SMEs.  \citeauthor{Bountouni2023AStudy} suggested that the recently released (April 2023) EU Cybersecurity Act~\cite{ECSME} lacks easily interpretable recommendations, and its provision requirement had no distinction between SMEs and large organisations.

In their study, \citeauthor{Chandra2020} findings suggest that SMEs must overcome many barriers, such as domain knowledge, solution complexity, infrastructure requirements, organisational changes, and financial costs~\cite{Chandra2020}. Furthermore, studies concluded that challenges were intensified in developing countries due to a lack of adequate technological infrastructure, lower levels of literacy, and the higher costs of acquiring new technologies~\cite{Kabanda2018, Chandra2020, Nasir2020}.

Two additional studies reported that the SMEs they studied insisted on implementing unsuitable solutions, which caused more problems than they solved~\cite{VanHaastrecht2021, Yudhiyati2021}. These failed implementations led to the inefficient use of resources (human, financial and silicon), loss of confidence and the further de-prioritisation of cybersecurity initiatives. 
A contrasting position emerged from \citeauthor{Tam2021}'s findings. They suggested that solutions have become suitable for SMEs. However, their deployment and operation remained unsuccessful due to SME employees' limited literacy and a different set of priorities~\cite{Tam2021}.  \citeauthor{VanHaastrecht2021} conclude that instead of creating new solutions, the industry should focus on tailoring existing ones to make them suitable for SMEs~\cite{VanHaastrecht2021}. However, it should be noted that empirical data did not support this latter conclusion.

\subsection{Literature overlooked SMEs' specific needs}

Many studies (n=34) claimed that the cybersecurity literature lacked a focus on SMEs. Literature gaps emerged in the introduction of 25 studies and were discussed in the conclusions of nine. Novel findings surfaced from two perspectives: areas requiring further exploration and deficiencies related to reporting standards. The discussion section delves into the latter perspective.

The introductions to six studies cited previous work that suggested existing recommendations were not actionable by SMEs either because they were theoretical, or too complex for the limited number of staff and SME cybersecurity personnel with low levels of cybersecurity literacy ~\cite{Carias2021, Rae2019, Ahmed2021, Ikuero2022, VanHaastrecht2021, Yudhiyati2021}. Additionally, one of the papers mentioned in its introduction that the literature remained impractical and lacked sufficient evidence to support the conclusions drawn~\cite{Ikuero2022}. 
In introducing their work, other studies highlighted gaps in the literature about topics including a lack of clarity regarding the implications of cybersecurity breaches~\cite{Durowoju2020}; data privacy management for SMEs~\cite{Jantti2020}; specific challenges faced by SMEs based in developing countries~\cite{Nasir2020, Yudhiyati2021}; a focus on Information Technology (IT) but not on IoT specific to SMEs~\cite{Yudhiyati2021}; inadequate attention to human factors in SME cybersecurity~\cite{Dykstra2020, Rae2019, Rae2020}; understanding the targeting of SMEs in supply-chain attacks and educating SMEs about such risks~\cite{Durowoju2020}; investigating cyber-risk management requirements and challenges for SMEs~\cite{Alahmari2020, Alahmari2021, Yudhiyati2021}; and, a lack of coverage on Information Security Management Frameworks (ISMF) in the context of SMEs~\cite{Perez-Gonzalez2019, Auyporn2023AStandard}. For two studies, the findings suggested that there was a lack of investigation on action taken by SMEs to protect themselves against cybersecurity risks~\cite{Berry2018AnThreats, Yudhiyati2021}.

As for novel findings, research that involved surveys with UK SMEs urged for existing solutions to offer more adaptability and lightweight versions that are sensitive to the constraints faced by SMEs~\cite{VanHaastrecht2021, Auyporn2023AStandard}. One of these studies also reported that SME leadership had little interest in participating in surveys, which limits research development in the field. For instance, low responses (<1\%) limited the possibility and representativeness of quantitative research~\cite{VanHaastrecht2021}.

A SR concluded that the recommendations provided were not actionable by SMEs~\cite{Bada2019}.
A survey of SMEs and large corporations reported an unmet need for a cybersecurity measurement framework specially focused on SMEs~\cite{Park2021}. 
Three empirical studies confirmed the need to investigate the specific needs and challenges of SMEs in developing countries~\cite{Alahmari2020, Kabanda2018, Alahmari2021}; a point that also emerged in the introductions of three papers~\cite{Nasir2020, Yudhiyati2021, Chidukwani2022}. 
Combined, all five studies implied that the existing literature focused predominantly on developed countries, and their findings were inapplicable to SMEs in developing countries. Research from \citeauthor{Chandra2020} suggested that challenges faced by SMEs in developing countries were intensified by limited access to funds, government support, cybersecurity literacy, and the cost of solutions~\cite{Chandra2020}. Section \ref{sec:developing} delves further into the geographical coverage of the literature.

\subsection{Low perception of cybersecurity risks}
\label{res:perception}

Perception and awareness may have overlapping meanings, and to mitigate this confusion, we employed the following definition. Lack of awareness is about not knowing a topic's existence. In contrast, low perception involves partially understanding its existence but not paying enough attention.

The theme \emph{low perception of cybersecurity risks} motivated research in 22 studies and emerged from novel findings in 18, being a total of 31 unique studies. This consistent emergence of drivers and conclusion classes (re)confirms the relevance of this theme in the literature.

This theme emerged from various distinct perspectives. 
In the first perspective, many studies suggested that SMEs did not pay attention to cybersecurity risks due to the misconception that they, as businesses, are too small and irrelevant to be targeted by cyberattackers. This argument was used to introduce 16 studies~\cite{Saban2021, Dykstra2020, Elezaj2019, Mitrofan2020, Ahmed2021, Kabanda2018, Alahmari2021, Young2020, Ogunyebi2018AnSector, Asprion2023, Franco2023CyberTEAInvestment, Jamil2024Human-centricBehaviours, Eybers2022InvestigatingSMEs, Wilson2023, Ponsard2020MethodologyBelgium, Ponsard2019SurveySMEs} and emerged from the findings of nine empirical studies~\cite{Saban2021, Dykstra2020, Mitrofan2020, Rae2020, Kabanda2018, Osborn2018, Chidukwani2022, Jamil2024Human-centricBehaviours, Wilson2023}. These findings suggested that SMEs believed attacks were concentrated on larger organisations and that their defences were adequate.
Research that interviewed SMEs highlighted that micro businesses perceived that they are not susceptible to cyber threats, or the threats are not capable of causing harm~\cite{Jamil2024Human-centricBehaviours}. 
Another interview showed that Australian SMEs believed they could protect themselves by limiting their online presence despite the significant business benefits of a greater online presence. 
\citeauthor{Wilson2023} report included a statement that nicely summarises all findings (and citations) supporting this perspective: `\emph{Despite their importance to the UK economy and their dependency upon I.T, our findings suggest that SMEs are still discounting their risk of cyberattack. This is despite knowing that if an attack were to occur the consequences could be significant. Perceptions of vulnerability appear to be tied to the notion that most cyberattacks are targeted. This is fuelled by SMEs' misconceptions about size, turnover, reputation rather than the indiscriminate approaches used by cybercriminals}'~\cite{Wilson2023}.

The second materialised perspective was that SMEs paid limited attention to cybersecurity because of a low perception of the benefits of defence~\cite{Rae2019}. Another (third) perspective was that SMEs overlooked cybersecurity because they underestimated the monetary value of their digital assets (data and infrastructure)~\cite{Ulrich2020}.
A fourth perspective surfaced in the introduction of three studies~\cite{Alahmari2021, Jantti2020, Yudhiyati2021}, which cited reports that SME leadership exhibited lower levels of perception of cybersecurity risk, which impacted upon their subordinates~\cite{Alahmari2021, Jantti2020, Yudhiyati2021}. Similar arguments emerged in the conclusions of nine studies, which suggested that low perceptions of risk amongst decision-makers result in the inadequate prioritisation of cybersecurity defences~\cite{Saban2021, Mitrofan2020, Kljucnikov2019, Alahmari2021, Heidt2019, Jantti2020, Yudhiyati2021, Jamil2024Human-centricBehaviours}. They recommended that cybersecurity initiatives should begin with top management; otherwise, employees would not prioritise it. Moreover, without leadership prioritisation, cybersecurity functions depended on small and often unrelated teams and were generally downplayed.
A fifth perspective materialised from a survey of SMEs in the UK.  In this study, it was reported that 75\% of organisations did not accurately perceive how (un)prepared their cybersecurity defences were. The same study indicated a sharply rising cost for SMEs to recover from cyberattacks. Yet, the perception gap remains significant compared with larger entities~\cite{Rae2020}.
A sixth perception was that SMEs perceived fewer cybersecurity risks when operating on cloud infrastructure. This discovery was observed in the conclusions of two studies~\cite{Saa2017, Rae2020}. Finally, the seventh perspective stemmed from \citeauthor{Ahmed2021}'s discoveries that the reasons behind the perception problem remain unknown in the literature and consequently require further research~\cite{Ahmed2021}.

\subsection{Overwhelmed, sparse or non-existent cybersecurity leadership}
\label{res:overwhelmed}

This topic emerged from 26 unique studies, where 16 reported original results, and 13 touched on this topic to introduce or motivate their research. It is worth noting that all findings surfaced from self-reported interviews, but there was no investigation testing these points.  All studies concurred that the absence of a cybersecurity unit posed a barrier to SMEs' cyber defences. Consequently, cybersecurity duties were assigned to other departments with other priorities and often had limited cybersecurity expertise.

This theme emerged from two angles. The first relates to cybersecurity functions being carried out by unrelated teams prioritising their primary activities, leaving security a step down on a long list of priorities. This notion emerged from the introduction of one study~\cite{Mayer2017} and the findings of two~\cite{Wilson2023, Mayer2017}. Building up to this argument, additional research indicated that allocating cybersecurity functions to leaders of unrelated domains resulted in overwhelmed managers accumulating too many functions. This finding was discussed in the introductions of three studies~\cite{Alahmari2020, Chandra2020, Loffler2021} and among findings of seven other reports~\cite{Alahmari2021, Barlette2019, Heidt2019, Heidt2018TheSecurity, Jantti2020, Kljucnikov2019, Saban2021}.  
Moreover, \citeauthor{Heidt2019}'s findings suggested the absence of dedicated cybersecurity leaders resulted in functions being picked up by leaders without adequate expertise~\cite{Heidt2019, Heidt2018TheSecurity}. Therefore, these managers required more time to learn and develop the function effectively.  They also indicated that the combination of additional time requirements and competing priorities led to the further deprioritisation of cybersecurity activities. The results of an interview went further and revealed that conflict of interest is one of the reasons for downplaying cybersecurity~\cite{Osborn2018}. To explain, researchers concluded that the effectiveness of cybersecurity relied on the ability of the organisation to restrict insecure practices. However, the decision to implement such restrictions is compromised when the same manager owns both the security control and the function or practice considered insecure. For example, the same manager is responsible for restricting access to sensitive data; at the same time, its team is the one accessing this data. Recommendations were concentrated on having dedicated and specialised cybersecurity functions. To add perspective to this notion, \citeauthor{Erdogan2023} surveyed 141 UK-based SMEs and reported that only a third (n=45) of organisations have employees dedicated to cybersecurity.

The second angle is about SMEs having dedicated cybersecurity personnel; however, they were too busy. This angle emerged from the findings of a single paper~\cite{Carias2020}. In that study, the authors indicated that the success of cybersecurity initiatives depended on, among other factors, leaders' ability to propagate awareness, commitment and engagement to cybersecurity with their subordinates. Consequently, busy, uninterested or non-proficient leaders were the cause of disregarded cybersecurity initiatives.  According to this same study, SME senior managers should drive and incentivise cybersecurity activities.

\subsection{Poor security operations}
\label{res:poorsec}

Nineteen studies suggested that SMEs have weaker cybersecurity defences than larger organisations. According to these studies, this is because SMEs often struggle to build robust and efficient security operations. Of these studies, 10 cited this theme to introduce their research, while 11 focused on investigating and presenting findings related to this issue.

To introduce their study, \citeauthor{Rawindaran2021} quoted previous research suggesting that SMEs put themselves at risk by deploying open-source software in an uncontrolled manner~\cite{Rawindaran2021}. 
Analogous discoveries were used to introduce five studies that SMEs were at risk due to uncontrolled deployments of IoT (Internet of Things) and personal devices (BYOD)~\cite{Heidenreich2019, Alahmari2021, Jantti2020, Ulrich2020, Yudhiyati2021}.
The findings from studies that involved interviews with SMEs supported this concern by suggesting that devices (IoT and BYOD) were deployed without effective cybersecurity posture assessment and monitoring, creating exploitable paths to corporate networks and sensitive data held by the organisations~\cite{Ncubukezi2020, Ikuero2022, Tam2021, Chidukwani2022}. 
Furthermore, one of these studies reported that SMEs used mobile phones as employees' primary computing devices~\cite{Ikuero2022}. 
The findings of another three studies indicated that SMEs struggled to maintain up-to-date inventories of their data and computing devices. This creates suboptimal cybersecurity as enterprises need a clear understanding of what they need to protect~\cite{Douchek2020, Carias2020, Yudhiyati2021}.

Two other empirical studies reported that SMEs used outdated software without the necessary controls to detect and update patches~\cite{Dykstra2020, Yudhiyati2021}, while \citeauthor{Douchek2020} found that SMEs stumble with cybersecurity because they fail to monitor the security posture of their digital assets (servers, systems and data)~\cite{Douchek2020}.

A case study of Peruvian and Mexican SMEs implied that their operational efforts were restricted to antivirus software tools, and that they did not have essential elements such as password policies or identity and access management. This posture resulted in at least one cybersecurity incident in the past 2 years~\cite{Diaz-Piraquive2023CybersecurityColombia}.

A literature review exploring the cybersecurity practices of  53 Italian SMEs' alignment with the NIST Cyber Security Framework (CSF) reported that only 31.5\% had a cybersecurity policy in place~\cite{Neri2024}. The same study also noted that SMEs deemed it unnecessary to establish a cybersecurity policy because they believed that employees already knew how to behave securely and that their systems already had sufficient levels of protection.

\subsection{Cloud adoption and managed services minimises cybersecurity challenges}
\label{res:cloud}

While our search did not explicitly focus on cybersecurity risks associated with the Cloud and managed services, this theme emerged from 14 studies. It was observed as a motivator in the introduction sections of nine studies and the results and conclusion sections of six. These studies generally found that SMEs perceive cloud adoption as an enabler in their cybersecurity journey.

For example, findings from an interview-based investigation of UK SMEs' cybersecurity preparedness~\cite{White2020} and a survey of cyber threat perception across Hungarian SMEs~\cite{Feher2020} suggest that SMEs perceived fewer cybersecurity risks when operating in the cloud. In both studies, respondents felt that the cloud provider took care of the risks. This same argument was also used to introduce another two studies~\cite{Dykstra2020, Ozkan2020}. At the same time, research suggested that while cloud adoption did address some of the risks, it introduced new ones, thereby shifting the cyber threat landscape. 
This issue was discussed in the introductory sections of three studies~\cite{Feher2020, Johannsen2020, Jantti2020}, and novel findings concerning this notion were presented in three others~\cite{Alahmari2020, Ahmed2021, Alahmari2021}. \citeauthor{Alahmari2020}, who authored two of these studies~\cite{Alahmari2020, Alahmari2021}, suggested that SMEs perceived fewer risks because: a) they misunderstood the shared responsibility model, mistakenly assuming that risks were addressed by their cloud provider; and, b) they could not see the new risks introduced by outsourcing their infrastructure.

An interview-based study, this time investigating the impacts of computing virtualisation on SMEs, reported that cloud adoption improved system availability and resulted in less frequent and shorter periods of outages~\cite{Ahmed2020}. Meanwhile, \citeauthor{Saa2017} cited previous research indicating that cloud adoption had increased SME access to solutions and technologies~\cite{Saa2017}. 
While none of the literature reviewed tested these hypotheses, a complementary view was brought by \citeauthor{Ahmed2020}'s conclusion that there was a reduced cost to SMEs associated with migrating their infrastructure and cybersecurity services to the cloud~\cite{Ahmed2020}. Equivalent statements were used to introduce another three studies~\cite{Mitrofan2020, McLilly2020, Carias2020}, and a partially contrasting point emerged in the introduction to one study~\cite{Yudhiyati2021}, which referred to previous research that indicated that in developing countries, cloud infrastructure was not affordable for SMEs. 
Along the same lines, one study reported that outsourced systems resulted in improved incident response capabilities, as they were handled by the service provider \cite{Ozkan2020}.  A complementary view was brought by a case study, which reported that SMEs presented higher cyber resilience after suffering an attack because this led to them subsequently outsourcing systems operations \cite{Carias2021}. Lastly, only one study reported that SMEs maximised resources and saved time through outsourcing~\cite{Carias2020}.

Moreover, other research suggests that SMEs resist cloud adoption~\cite{Johannsen2020, Yudhiyati2021}. One of these studies cited previous research showing that German SMEs have not yet adopted cloud-based technologies because of concerns about the cybersecurity readiness of these services~\cite{Johannsen2020}. In their survey of Middle Eastern SMEs' cybersecurity posture, \citeauthor{Ahmed2021} reported a similar adoption pattern but for different reasons. They found that 50\% of respondents were still planning or not interested in using cloud computing~\cite{Ahmed2021}. 
Further, a case study of Saudi based SMEs reported that SMEs were concerned that service providers would not keep the pace with the SME's development \cite{Almubayedh2018}.

\subsection{Legislation and informative content improve cybersecurity}
\label{res:legal}
The notion that standards and regulations positively affect cybersecurity awareness and measures surfaced in the introduction of five studies and the findings of 7, a total of 11 distinct studies. 

To motivate their research, five studies stressed the importance of region- and industry-specific legislation such as the General Data Protection Regulation (GDPR) and the Health Insurance Portability and Accountability Act (HIPAA) to foster cybersecurity initiatives among SMEs~\cite{Elezaj2019, Bada2019, Ikuero2022, Gafni2019, Saban2021}. According to the papers cited, the benefits of these regulations were more about their informative aspects, as opposed to punitive directives. The knowledge of their existence alone was enough to stimulate SMEs to consult with cybersecurity providers, which led to increased cybersecurity awareness and measures.  Nevertheless, we encountered no evidence of this notion being soundly tested. The closest validation emerged from a study in which 25 Belgium-based cybersecurity experts were interviewed~\cite{Ponsard2020MethodologyBelgium}. Their findings suggested that while GDPR remains an excellent incentive for raising awareness, it resulted in GDPR-consulting companies positioning themselves as cybersecurity specialists. According to the authors, while the activities prescribed by GDPR represent a fraction of the cybersecurity programme,  SMEs believed it was enough to protect them against all cyber threats~\cite{Ponsard2020MethodologyBelgium}. In another study, which this time involved interviews with  68 owners of UK- and Saudi-Arabia-based SMEs, it was pointed out that regulatory frameworks had a (positive effect) chain reaction. This started by raising cybersecurity awareness, followed by increased cybersecurity literacy, which resulted in more resources being allocated to cybersecurity initiatives and consequently improved cyber resilience overall~\cite{Rawindaran2023EnhancingKingdom}.

Despite the benefits discussed, the findings of survey research concerned with the cybersecurity measures and challenges encountered by SMEs from Indonesia~\cite{Yudhiyati2021} and the Czech \& Slovak Republics~\cite{Douchek2020} suggest that regulations could be better tailored to SMEs, for example, taking into account the limited resources available to these organisations.

In their study's motivation, \citeauthor{Rae2019} quoted government initiatives that promoted cybersecurity standards for SMEs. These initiatives provided SMEs with the opportunity to offer services in the public sector, on the condition that they establish at least baseline cybersecurity practices. As an example of how governments are incentivising this, the UK government set a target for 2022 to procure at least 33\% of their contracts from SMEs who adhere to the Cyber Essentials standard~\cite{Rae2019}.
A contrasting position emerged from a survey (conclusion) of SMEs in developing countries. Their findings suggested compliance-driven efforts can negatively impact SME cybersecurity as baseline requirements may not be sufficient to protect the organisation. Still, they overshadow the need for more robust practices~\cite{Kabanda2018}.

\subsection{Remaining themes}
\label{res:remaining}

So far, this section has explored the most commonly discussed themes in the literature, from which two themes remain to be addressed: the \emph{rise in cyberattacks against SMEs} and the \emph{increasing financial losses due to cyberattacks}. 
The discussion of these themes was mainly observed in the introduction sections of the reviewed papers, but they barely emerged from the research conducted, and when they did, this was not based on empirical findings. 
The former theme was touched upon in the introduction section of 41 studies, but no investigation was conducted to measure the evolution of cyberattacks over time.  This theme emerged from studies citing industry reports (discussing an increase in financial losses) and previous research stating a point-in-time number of cybersecurity incidents; consequently, more investigation is required to support the concept that attacks are actually on the rise. 
The latter theme, \emph{increasing financial losses due to cyberattacks}, was touched on to motivate the research reported in 17 studies and the conclusion section of two, though these studies did not primarily investigate this subject. To explain, the first instance surfaced from a literature review~\cite{VanHaastrecht2021}. The second instance emerged from the following statement (conclusion) of a study investigating behavioural assessment models. 
`The average cost for SMEs suffering a cyberattack has dramatically risen, yet the perception and awareness gap is still significant compared to larger businesses'~\cite{Rae2020}.

\section{Discussion}
\label{sec:discussion}

In this section, we address the research questions --- presented in the introduction --- based on the findings from the review.  We then discuss the current state of research on SME cybersecurity and provide a view of its maturity, considerations of encountered reporting standards and research gaps. Lastly, we discuss how knowledge has been perpetuated over time without being retested.

\subsection{Cybersecurity threats experienced by SMEs}

The first question, \emph{`RQ1: What are the cybersecurity threats experienced by SMEs?'}, was not fully answered in the reviewed literature. Instead of concentrating on threats, studies often focused on the challenges SMEs encounter in establishing cybersecurity defences. They suggested that SMEs were unaware of the threats and that existing solutions did not address the needs of SMEs. Yet, none investigated or articulated the threats faced by SMEs. Nor did they consider if and how SMEs' threats differ from those of larger organisations. Similarly, 41 studies introduced their research stating that cyberattacks against SMEs were rising; however, no research was found to test the notion that the frequency of cyberattacks targeting SMEs is indeed growing. While reviewing studies, we identified datasets such as the UK Cybersecurity Breaches Survey~\cite{UKCyberBreach2023} and the Verizon~\cite{Verizon2022} Data Breach Investigations Reports, which provide raw (primary) data regarding historical occurrences of cyberattacks. However, while many studies (n=8) cited their existence, none used the data they contain to test whether there was an increasing frequency of cyberattacks (see theme \emph{Rising cyberattacks against SME} in Table~\ref{tab:themes-tb}).
Unfortunately, based on the literature reviewed, it remains unclear whether cyberattacks against SMEs are on the rise.  Future research should therefore explicitly test this.  In doing so, such research should consider that SMEs are known to under-report cyberattacks~\cite{Gafni2019, Ikuero2022, White2020, McLilly2020, Chidukwani2022, Neri2024} and hence that any estimates are likely to be conservative if they are based on attacks reported to the authorities.  Attention might also be given as to whether reporting rates vary over time.  The analysis of data collected for the UK Breaches Survey, which does not rely on attacks reported to the authorities, should, however, mitigate these concerns.

The search of the grey literature identified many industry reports that focused on data breaches and often attempted to forecast upcoming threats. However, it is essential to note that these reports primarily concentrated on more prominent organisations and made no distinction between the threats faced by these organisations and SMEs. 
The grey literature also highlighted a class of services called \emph{threat intelligence}, whereby companies monitor cyberattacks in various industries and regions. This information is then aggregated to produce reports (behind paywalls) on attackers' motivations and techniques. This type of service is well described by the Gartner research institute: `\emph{Threat intelligence is evidence-based knowledge (e.g., context, mechanisms, indicators, implications and action-oriented advice) about existing or emerging menaces or hazards to assets}'~\cite{Findings2016}. 
To our surprise, the systematic search did not identify any studies discussing threat intelligence for SMEs. However, a more specific search that focused on threat intelligence and SMEs identified a single study that advocated for an approach to threat intelligence tailored to SMEs~\cite{VanHaastrecht2021Threat}.

Despite almost half of the literature claiming that SMEs are unaware of the cybersecurity threats they face, no study was found to enumerate these threats. We were also surprised that we did not find a taxonomy of threats and risks.  This situation begs the question of how SMEs could be aware of the threats if no research has aimed to identify them. In an attempt to answer this question, we carried out an ad hoc search that revealed the existence of well-established cyber threat taxonomies~\cite{JRCTaxonomy, CarnegieTaxonomy, Rabitti2025}.

To conclude, further research is required to identify SMEs' threats and risks, compare them to those faced by large organisations, and develop an SME-focused taxonomy of threats.

\subsection{SME awareness of cyber threats}

Regarding the second research question, \emph{`RQ2: What awareness do SMEs have about the threats they face?'}, while the literature offered limited insights on threats, a substantial amount of research touches on SMEs' levels of awareness of the threats they might face and the need to invest in reducing their risk.  
 \emph{\hyperref[res:awareness]{Lack of awareness}} was the most discussed theme in the literature; it was present in 57 studies. In all instances, studies suggested that this issue was widespread from employees to leadership. Of these, nine studies reported a positive relationship between the levels of awareness expressed by employees of SMEs and their behaviour in terms of cybersecurity~\cite{Mokwetli2018, Alahmari2020, Raineri2020, Kljucnikov2019, Heidt2019, Pickering2021, Perez-Gonzalez2019, VanHaastrecht2021, Saban2021}.

Despite most studies agreeing that levels of awareness were inadequate, five studies presented a contradictory picture, suggesting that SMEs had awareness but that their approach to cybersecurity fell short for reasons including resource constraints and limited cybersecurity literacy~\cite{Dykstra2020, Pickering2021, Osborn2018, Carias2020, Ulrich2020}. While this alternative perspective was explicitly discussed in only five studies, this concept is implicitly supported by a more significant number of studies that examined issues associated with the themes \emph{\hyperref[res:literacy]{limited cybersecurity literacy}} and \emph{\hyperref[res:underfunded]{under-funded and resource-constrained cybersecurity programmes}} (discussed in sections~\ref{res:literacy} and~\ref{res:underfunded}).
To summarise, nearly half (43\%) of the studies reviewed discussed inadequate levels of awareness as a deterrent to cybersecurity in the context of SMEs. At the same time, a small number (n=5) of studies implied that limited literacy and resource constraints were the underlying problem (instead of awareness). We believe further investigation is required to understand the correlation between awareness, literacy and resource availability issues. These three themes were the most frequently discussed subjects in the literature. Yet, their relationship is poorly understood, and the volume of empirical research that tests these ideas is more limited than the volume of studies that discuss them. This situation is further addressed in section~\ref{disc:sythesis}.

\subsection{Practised cybersecurity measures}

The third question, namely \emph{`RQ3: What is the uptake of cybersecurity controls by SMEs?'}, is yet to be addressed. Most of the literature (102 studies) employed interviews and surveys primarily to identify challenges rather than practised measures. Consequently, these reports emphasised what SMEs lacked, but not what they had done. No research investigated existing controls and their effectiveness.
Nevertheless, we could still identify insights regarding SMEs' controls, at least inefficient ones.  Studies that explored the themes of \emph{\hyperref[res:poorsec]{poor security operations}} and \emph{\hyperref[res:overwhelmed]{overwhelmed, sparse or non-existent cybersecurity leadership}} (discussed in sections~\ref{res:poorsec} and~\ref{res:overwhelmed}) highlighted inefficient controls such as the uncontrolled deployment of open-source software and unmanaged anti-virus software. These same studies also reported how these (scarce) initiatives lacked dedicated and specialised ownership.

We would be remiss not to question the reliability of findings in this context. As previously mentioned, studies relied heavily on SME personnel to provide data (interviews and surveys) on employed measures and their efficiency. However, \emph{lack of awareness} and \emph{limited literacy} (discussed in sections~\ref{res:awareness} and~\ref{res:literacy}) emerged among the top three most frequent themes, which indicates that the same personnel were also considered unaware and lacking literacy in terms of cybersecurity dangers and required controls.  How can we rely on their responses and, at the same time, claim they lack literacy in cybersecurity?

Consequently, it is evident that there is a need for further investigation into the uptake of cybersecurity controls amongst SMEs and if and how this differs from larger organisations. Furthermore, the data collection method must not only rely on SME personnel (self-reported) or reported cyberattacks, as these are often not reported (see previous discussion).

\subsection{Availability of cybersecurity solutions}

About the fourth and fifth research questions, \emph{`RQ4: Which cybersecurity frameworks apply to or can be tailored to SMEs?'} and \emph{`RQ5: What challenges do SMEs face in adhering to existing cybersecurity frameworks and solutions?'} respectively, we found 14 studies that investigated how existing solutions catered to SMEs' needs. 
These studies unanimously concluded that existing solutions remain unfit. We found no research that identified the existing solutions or provided a taxonomy on classes of solutions. Instead, we encountered a series of generic statements and sometimes theoretical assessments of how existing solutions were unsuitable for SMEs. We expected these investigations to provide more details on unmet requirements or shortcomings of existing solutions. 
Rather, findings were restricted to comments such as existing solutions being unsuitable for SMEs due to resource and knowledge constraints. 
For instance, these same studies reported that SMEs struggled to implement solutions due to the requirement for highly specialised domain knowledge regarding infrastructure, organisational changes, solution complexity and increased financial costs.
They also suggested that these challenges were intensified in developing countries due to further limitations associated with cybersecurity literacy and financial resources.
To conclude, \emph{`RQ4'} is partially answered and \emph{`RQ5'} remains unaddressed. We found this result surprising, considering the existence of numerous frameworks intended for small organisations such as the UK Government Cyber Essentials~\cite{CyberEssentials} and the lite version of the Cloud Control Matrix from the Cyber Security Alliance~\cite{CSA-CCM-lite}, both designed for and commonly known to address SMEs issues. Yet, no study was found that had tested their suitability.

It is essential to emphasise that  34 studies mentioned that existing solutions do not serve SMEs' needs, of which 28 only touched on this theme to introduce their research. Moreover, 64 studies proposed custom frameworks, and while they focused on validating their solution, the investigation of existing ones was superficial. Proposed frameworks were either entirely theoretical or had not progressed past a limited initial feasibility study. No usable deliverables (documentation, software, source code, etc.) were made available, which severely limits the ability of others to take these ideas forward.

More investigation is required to provide actionable information on how existing solutions must adapt to serve SME needs. The current literature provides examples of individual initiatives, but these were scattered and without cumulative progress.

\subsection{Overall model of SME cybersecurity dependencies} 
\label{disc:sythesis}

\begin{figure}[!htbp]
    \centering
    \includegraphics[width=0.47\textwidth]{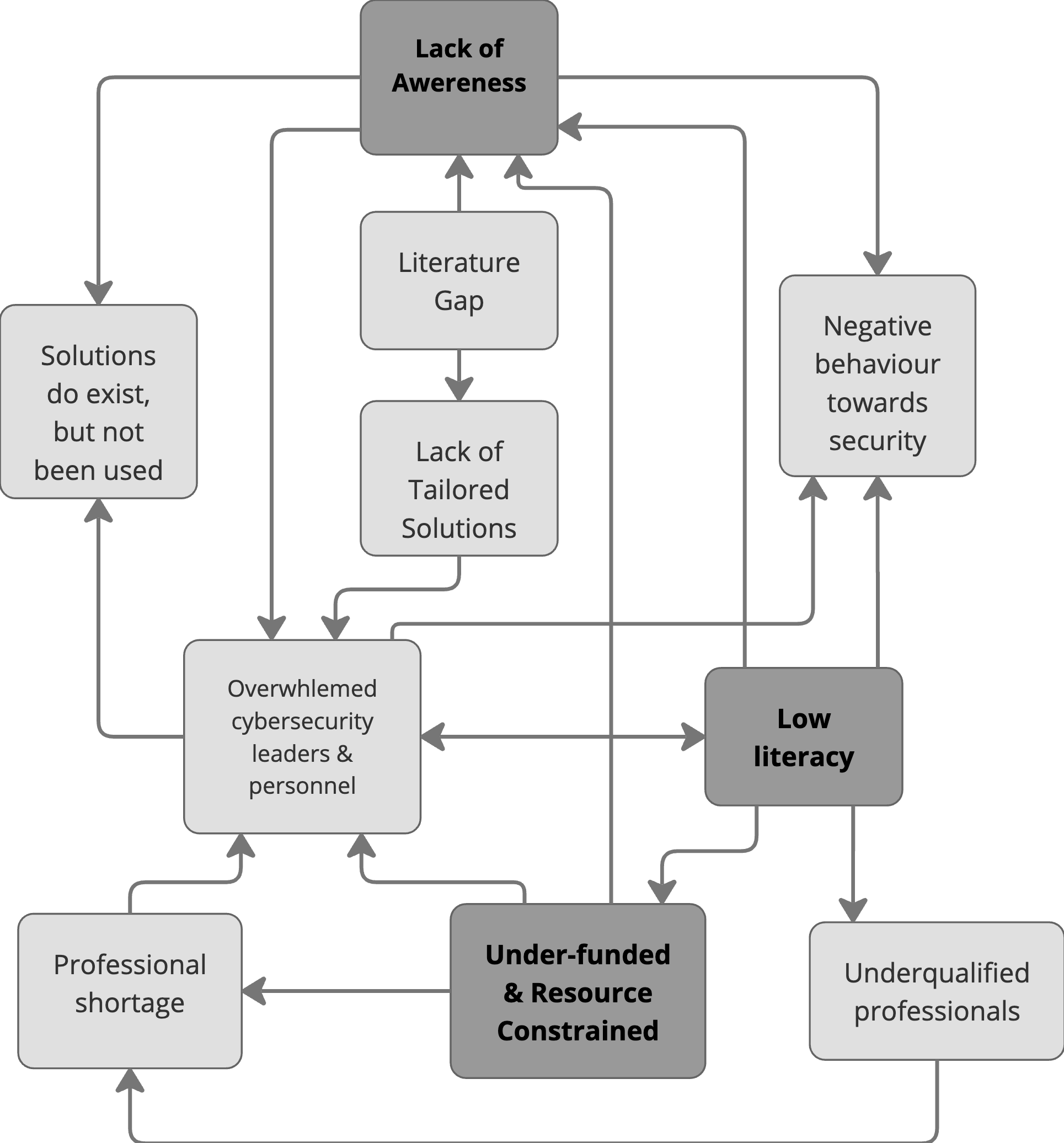}
    \caption{Potential dependencies between the themes that emerged.}
    \label{fig:correlation}
    \figalttext{Directed graph linking ten themes that emerged in the review. Three central themes are highlighted in darker shading: lack of awareness (top), low literacy (middle right), and under-funded and resource-constrained (bottom). Arrows indicate low literacy and under-funding feeding into lack of awareness, into overwhelmed cybersecurity leaders and personnel, and into negative behaviour towards security; the literature gap leads to a lack of tailored solutions; professional shortage and underqualified professionals connect back to under-funding and low literacy.}
\end{figure}

While the result section examined emerged themes in isolation, they are not mutually exclusive and were often discussed together in particular studies. Even though no research was found to explore causality among themes, we identified potential relationships. For instance, themes discussed in the study conclusion sections were often discussed in causal terms. These connections are visualised in Fig.~\ref{fig:correlation}.

A substantial volume of research (n=30) concluded that a \emph{lack of awareness} was the main reason for SMEs having ineffective cybersecurity initiatives. The majority of these studies measured awareness levels, and not many focused on the causes of the low level of awareness. For instance, three studies indicated that \emph{limited literacy in cybersecurity} resulted in lower levels of awareness~\cite{Pickering2023ITechnology, Neri2024, Rawindaran2023EnhancingKingdom}. Despite this not being explicitly stated in their findings, we would argue that more of the studies reviewed presented data that would support this conclusion. Our interpretation is that the \emph{lack of awareness} issue resulted from \emph{limited literacy in cybersecurity} and \emph{constrained resources}. Some studies suggested that the \emph{constrained resources} issue itself appears to be the result of \emph{limited literacy}, as the necessary resources did exist, but they were not allocated to cybersecurity~\cite{Heidt2019, Mitrofan2020, Barlette2019}. We recognise that the issue of resourcing is complex and depends on many other factors, such as government support, subsidies, skilled personnel availability, the geopolitical region, and the industry the SME operates within, which is further discussed below.

As a consequence of the relationship between the \emph{lack of awareness}, \emph{limited literacy}, and \emph{constrained resources} being poorly understood in the literature, research and interventions may have been misdirected by approaching these issues in isolation. For example, studies have investigated interventions to raise awareness among SME personnel, without considering whether resources (staff, tooling, time and finances) are available to allow them to address shortcomings in cybersecurity, should attempts at raising awareness be effective. 
Such a situation could lead to an increase in anxiety as staff are more aware but not supported in solving the newfound issues.
More generally, raising awareness alone may be insufficient (and too simplistic) in the absence of adequate provisions regarding education, sufficient time and investment. Being aware of risks is one thing; knowing how and having the means to address them is another. Research has typically focused on the former, but without the latter, SME cybersecurity initiatives are likely to be unsuccessful.

Considering the aforementioned points, here we articulate a number of future research questions: Does raising awareness alone lead to improved cybersecurity? What is required to translate awareness into action? To what extent do resource constraints impact awareness? For instance, studies suggested that SME employees were overwhelmed and did not have adequate time to perform cybersecurity functions. Does giving them more time or fewer tasks improve cybersecurity performance in these SMEs? Can those involved learn from project management resource allocation? If more time is not the solution, how can we provide more structure to improve security task performance? Lastly, is limited literacy the fundamental shortcoming of SME personnel?  Can we state that awareness issues have resulted from a poor understanding of cybersecurity threats and their consequences, or does something else explain it?

\subsection{Developed versus developing countries}
\label{sec:developing}

\begin{figure*}[!htbp]
    \centering
    \includegraphics[width=\linewidth]{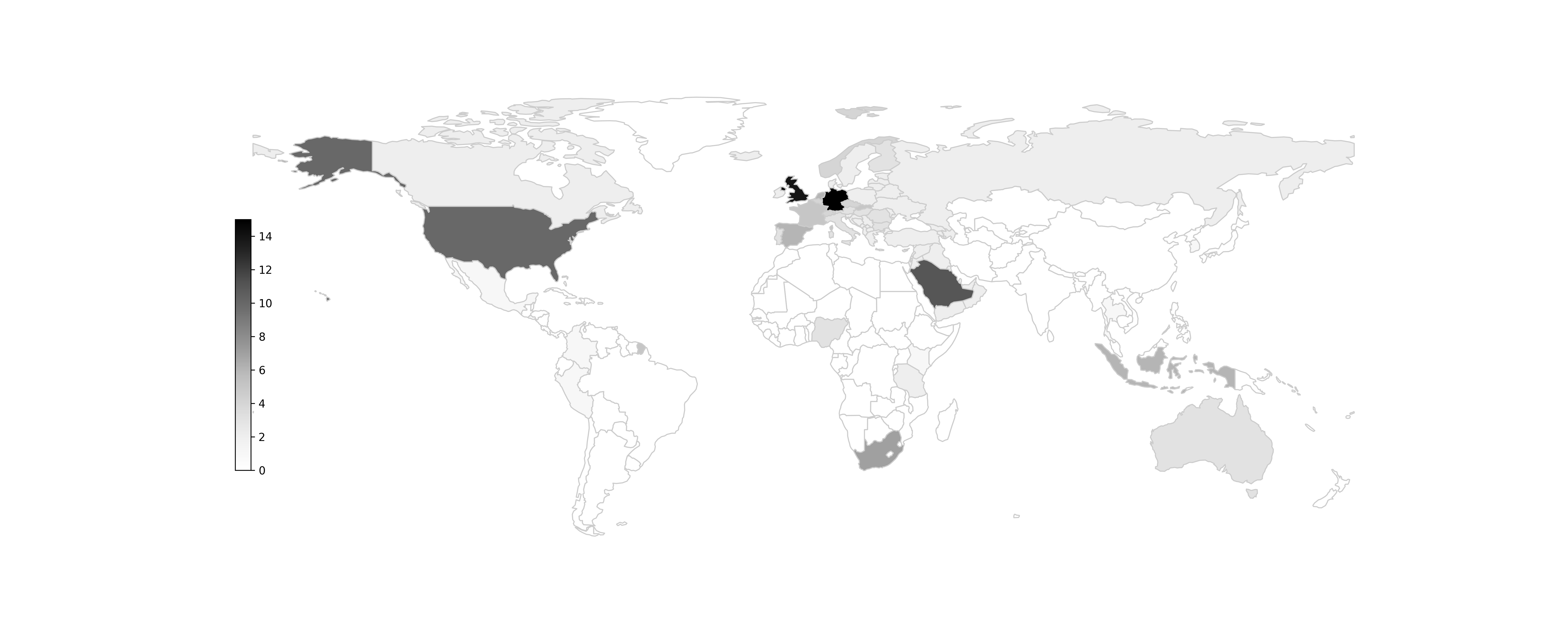}
    \caption{A heat map gauging the geographic region covered by reviewed literature. This data is available at (\url{https://osf.io/ps7xy/}).}
    \label{fig:geodistrib}
    \figalttext{World map shaded by the number of reviewed studies covering each country, on a grey-scale legend from 0 to 14. The darkest shading falls on Germany and the United Kingdom (nine studies each), followed by Saudi Arabia (seven) and Indonesia (five); the United States, several European countries and parts of Africa and South-East Asia show lighter coverage. Central and South America remain uncovered.}
\end{figure*}

Many studies claimed that the literature was limited to SMEs situated in developed countries~\cite{Alahmari2020, Kabanda2018, Alahmari2021, Nasir2020, Yudhiyati2021}. To examine this, we coded studies based on the geographic coverage of the SMEs studied, and our results suggest a different conclusion. That is, as shown in Fig.~\ref{fig:geodistrib}, the reviewed literature covered 99 countries, including those in Africa, Asia, Europe, and Oceania developing countries, whereas American (Central and South) SMEs remained unexplored. The highest coverage was in Germany and the UK (nine studies each), followed by Saudi Arabia (seven studies) and Indonesia (five studies).

Our data revealed substantial research on developing countries, thereby rejecting the idea that studies are limited to developed nations. Nonetheless, certain developing regions, including South America, have yet to be investigated in the literature. To objectively compare these regions, research must account for their unique geopolitical and economic characteristics, such as wars, professional education, specialised professionals' availability and solutions' affordability. In undertaking such research, it is essential to account for the disparities in currency exchange rates between developing and developed countries. It intensifies the affordability issue, as solutions are often exported from developed regions. The section~\ref{res:underfunded} provides further details on challenges faced by SME based in developing countries.

\subsection{Defining SMEs}
\label{disc:def} 

We did not adopt a particular definition of SMEs for this review, and as such did not exclude any studies based on the definitions employed by study authors.  However, it is important to note that we encountered a range of definitions of what SMEs are across studies. Most authors relied on definitions provided by local governments and regional entities; however, they differed from one country to another. 
For instance, the Saudi Arabia Small Business Administration (SBA) and Census Bureau defined SMEs as any organisation with less than 500 employees~\cite{Almubayedh2018}. At the same time, the South African Business Act classified an SME as any organisation with up to 200 employees, with no more than ZAR 10 million invested in assets and a turnover of up to ZAR 64 million, equivalent to 0.54 and 3.3 million dollars, respectively~\cite{Mokwetli2018}. In comparison, the US Business Association categorised SMEs as organisations with up to 500 employees and a turnover between 7 and 250 million dollars, depending on the industry~\cite{USSME}. The European Commission~\cite{ECSME} and the UK~\cite{UKSME, UKSMESTAT} government categorise organisations with up to 10 employees or 2 million euros as micro-businesses, while those with 50 employees or 10 million euros turnover and 250 employees or 50 million euros are classified as small and medium enterprises, respectively. More global definitions were encountered, unfortunately, with no evidence of use. For example, the United Nations~\cite{NATOSMEdef}, IMF (International Monetary Fund)~\cite{IMFSME}, and the OECD (Organisation for Economic Cooperation and Development)~\cite{OCED-SME-DigTrans} defined micro organisations as those with up to 9 employees, while those with 49 and 249 employees were considered small and medium, respectively.
These are just examples of how established definitions varied. In addition, we discovered studies that applied their own definitions. For instance, research investigating small German organisations defined SMEs as businesses with up to 100 employees~\cite{Ulrich2020}.  

Differences in the definition are important beyond conceptual concerns. For example, when considering financial constraints, these are likely to vary for businesses with a turnover of USD 250 million and those with EUR 50 million or ZAR 64 (USD 3.3) million. The same applies to the number of employees.

\subsection{Variable definition of cybersecurity solutions}

Cybersecurity is a diverse domain, and solutions developed to address different problems are invariably grouped under the term `cybersecurity solutions'. This generalisation creates the false impression that studies were referring to the same class of solutions when those they have in mind could be quite different. For instance, many studies (n=24) suggested that existing `cybersecurity solutions' were unsuitable for SMEs, while others (n=23) proposed new `cybersecurity solutions'. While they all employed the same terminology, they touched on entirely different classes of solutions.
For example, some discussed information security management and risk management solutions~\cite{Ozkan2019, Ahmed2021, Alahmari2020, Alahmari2021}; others cybersecurity maturity measurement frameworks~\cite{Johannsen2020, Ozkan2020}; others cybersecurity awareness assessment~\cite{Lejaka2019}; others cybersecurity resilience frameworks~\cite{Carias2021}; others cybersecurity assurance and self-measurement framework~\cite{Carias2021, Johannsen2020, Heidenreich2019}; while others discussed automation and the standardisation of cybersecurity incident response capabilities~\cite{McLilly2020}. 

Despite the generalisation and the unrelated usage of the same terminology, studies unanimously concluded that existing solutions are still unsuitable for SMEs. \citeauthor{Osborn2018} suggested this was because the literature and industry mistakenly assumed that large and small organisations always shared the same requirements~\cite{Osborn2018}. However, in our view, it is likely that the cybersecurity industry and researchers alike overstate the impact of new technologies and understate the effort required to tailor them to smaller businesses.

\subsection{Standards of reporting in the literature}
\label{disc:reporting}

Variability in what and how it was reported posed one of the biggest challenges to extracting findings for this review (see also~\cite{Ikuero2022, Perez-Gonzalez2019, Tam2021}). For example, more than three-quarters (n=102) of the studies employed interviews and surveys in their research.  However, these studies frequently failed to report critical information, such as the approach used to sample SMEs (e.g. was random sampling employed), characteristics of the employees interviewed (e.g. role and tenure), and the data collection method used (e.g. structured or open questionnaires). Without such details, it becomes difficult (or impossible) to replicate these studies or assess the validity and generalisability of their findings.  This contrasts with more mature domains, such as the medical sciences, which have developed formal standards for reporting findings from empirical studies. For example, the Consolidated Standards of Reporting Trials (CONSORT) defines a minimally acceptable standard for studies employing medical trials~\cite{Eldridge2016}.
CONSORT was established in 1995 and has evolved over the years. The current version (2010) has been adopted by the leading journals of medical science~\cite{GillC2011}.

In the absence of a dedicated reporting standard in the cybersecurity field, we adopted some aspects of CONSORT to inform the inclusion criteria applied for this review. For example, studies were excluded if they did not comprehensively describe the research methods. As outlined in Fig.~\ref{fig:prisma}, of the 257 ($414-157$) studies assessed for eligibility, 125 ($113+12$) were excluded due to inadequate reporting standards.
This substantial number (49\%) of exclusions highlights the necessity for reporting standards in the field. Others have made similar observations. For example, researchers have raised problems with unbalanced samples, missing data points, untested findings and variable reporting standards~\cite{Ikuero2022, Perez-Gonzalez2019, Tam2021}. For the field to advance, this needs to change.

\subsection{Lack of progress in SME research}
\label{disc:lackprogress}

The approach we took of distinguishing between novel contributions in the literature (conclusions) and discussions of prior literature that motivated a study (drivers) allowed us to evaluate the progress made on these themes over time. 
This differentiation revealed that recent research has introduced only a few novel findings. However, authors repeat and propagate many early discoveries without empirically replicating them in modern environments or offering new insights. 
Unfortunately, and perhaps ironically, many empirical studies came to the same conclusions as existing research, but authors were often seemingly unaware of previous work on the topic. 

A further concern relates to the representativeness of the evidence base, which is affected by three distinct but inter-related issues.
First, the field is dominated by the use of qualitative methods: 71\% of the reviewed studies employed qualitative approaches (see section ~\ref{sec:demog}). Qualitative research is valuable for in-depth exploration of under-studied fields, and is particularly appropriate given the nascent state of SME cybersecurity research. However, while such studies can provide rich, contextual understanding they are not intended to produce statistically representative conclusions. Consequently, the relative shortage of quantitative research means that few findings have been tested at scale or across diverse samples of SMEs.
Second, the sample sizes used in previous research have frequently been small. For example, of the studies that reported a numeric sample, 40\% (n=38) involved fewer than 25 participants. Small samples limit the diversity of perspectives captured in qualitative work, and reduce statistical power in the few quantitative studies conducted, making it difficult to draw robust or generalisable conclusions from either approach.
Third, any study that relies on voluntary participation, whether qualitative or quantitative, large or small, is susceptible to volunteer bias (also known as self-selection bias). That is, SME personnel who agree to participate in the research may be quite different to those who do not.  For example, they may be more likely to be those who already possess awareness of, or who are invested in cybersecurity. This means that findings regarding levels of awareness, risk perception and adopted controls may systematically differ to the cybersecurity posture of the broader SME population they are intended to represent. The extent of this problem is difficult to assess, as the vast majority of studies did not report participation or response rates. Of the 102 studies that employed surveys or interviews, only 19 reported the response rate achieved.  This was typically <3\%, with the exception of one study by~\citeauthor{Auyporn2023AStandard} that reported a 52\% response rate.  As pointed out by \citeauthor{VanHaastrecht2021}: "\emph{Low responses (<1\%) limited the possibility and representativeness of quantitative research}" \cite{VanHaastrecht2021}. The absence of transparency regarding non-response makes it difficult to evaluate the degree to which volunteer bias may have affected individual studies, and by extension, the conclusions drawn from the literature as a whole.
These three issues compound one another: a field dominated by qualitative methods with small, self-selected samples offers limited means to assess how representative its findings are. We therefore urge researchers to broaden the evidence base through the use of both qualitative and quantitative approaches, and the use of more representative sampling strategies that explicitly draw upon sampling frames, and employ some form of random sampling.

Beyond the issue of representativeness, a few additional factors may explain the limited progress observed.
First, researchers often appeared to conduct superficial literature reviews, neglecting exploration of foundational research that could have informed their investigations. This became evident from the fact that the driver and conclusion themes (on the same subject) often emerged in different groups of studies (see the intersection $\cap$ column on Table~\ref{tab:themes-tb}).
Second, the absence of standardised research methodologies (and reporting standards) in the field of cybersecurity has created space for opportunistic practices (e.g. interviews with small numbers of SMEs that were known to the researchers as opposed to the random sampling of SMEs using a sampling frame), potentially overshadowing the use of more rigorous research methods that are the expectation in more established fields of enquiry. 
Finally, in addition to the variable reporting standards discussed in section~\ref{disc:reporting}, the validity and reliability of some cybersecurity literature have been questioned by other researchers. For instance, \citeauthor{Gross2021} assessed the quality of 114 quantitative studies related to cybersecurity (from 2006 to 2016). Gross suggested that two-thirds of the papers reported incomplete data, and about one-quarter presented findings based on erroneous calculations and inconsistent data. This research also concluded that reliability decreased over time, with there being more errors in newer studies~\cite{Gross2021}. In comparison, there was not a single instance of such errors in studies published in the same period by the Journal of Media Psychology (JMP).

Issues with the standard of (SME) cybersecurity literature would appear to be a more widespread issue, and solving it is not easy. It requires coordinated initiatives and effort from researchers, reviewers, funders and stakeholders. Our recommendation starts with the need for a definition/adoption of sound and rigorous research and reporting standards, as have been established in other fields (e.g. medicine~\cite{Eldridge2016, Moher2016, Higgins2019}), followed by more stringent review processes before publication.

\subsection{Limitations}
As with any research, this review is not without limitations. For instance, we only considered studies written in English, which could introduce a language bias and lead to underrepresentation of specific regions, especially developing countries. 
To ensure that the information we included was not outdated, we excluded studies published before January 2017. Doing so may have missed important information that was not subsequently discussed in the literature. Similarly, we may have missed information by excluding studies that did not explicitly describe their research methods. Nonetheless, we accepted this risk in favour of relying on findings from studies in which it was clear how the data were collected and analysed (a rather basic requirement). 
Lastly, qualitative research inherently has an element of subjectivity. This research was carried out by three authors who met regularly to review and discuss the findings.  Efforts were made to maintain objectivity, but biases can never be eliminated---only acknowledged and mitigated.

\section{Conclusion}

Despite SMEs being the backbone of the global economy, cybersecurity efforts remain focused on larger organisations. 
Nonetheless, there is a growing number of initiatives and research projects concerning SME cybersecurity. Here, we provide a SR of recent literature, applying qualitative methods to identify and synthesise the existing knowledge. We differentiated between knowledge that is perpetuated throughout the literature without new empirical findings and those findings that are based on the generation of new evidence. We discussed the most frequently emerging themes and their potential relationship. Our discussion included observations and recommendations concerning the variability of reporting standards across studies, variability in the reliability of the literature and a finding that some studies would report empirical work without being aware of previous work that had examined the same or similar issues.

Notwithstanding our reservations about the quality of some of the literature, three significant barriers to cybersecurity for SMEs emerged: a lack of awareness of threats faced, resource constraints (human, digital and financial) and limited cybersecurity literacy among SME personnel.
While there appears to be a \mbox{cause-and-consequence} relationship between these factors, research is required to better understand the mechanisms through which they affect each other.
We hope that the findings of this review will inform the future research agenda and help industry and governments deliver initiatives that are tailored to SME needs.

\subsection{Future work}
Our analysis suggested that limited cybersecurity literacy functions as a root cause that contributes to both an awareness deficit and the misallocation of resources (section~\ref{disc:sythesis}). If this relationship holds under empirical scrutiny, interventions targeting literacy, rather than awareness in isolation, are likely to yield sustainable improvements. However, the field currently lacks both a theoretical framework for what constitutes adequate cybersecurity literacy in the SME context and structured models for translating literacy into cyber resilience improvements. Future research should therefore pursue two complementary objectives. The first is the development and validation of a cybersecurity literacy framework that accounts for SME heterogeneity, distinguishing between the resources required by companies of different sizes (small, micro and medium) and the proficiency levels across knowledge domains (e.g. threat identification, risk assessment, incident response, regulatory compliance).  For instance, an organisation with 200 employees is more likely to have a specialised cybersecurity function than a business with fewer than 10 employees.
The second is empirical research intended to test if cybersecurity literacy is indeed the root cause of a lack of awareness and restrictions in resource allocation towards cybersecurity initiatives. This in turn, would inform the development of government and industry-led policy mechanisms, regulatory frameworks and industry led-products and services.

\section*{Author contributions}
Carlos Rombaldo Junior (Writing--original draft), Ingolf Becker (Supervision, Writing--review \& editing), Shane D. Johnson (Supervision, Writing--review \& editing).

\section*{Conflicts of interest}
None declared.

\section*{Funding}
None declared.

\begin{appendices}
\renewcommand{\thesection}{\arabic{section}}
\setcounter{table}{0}
\renewcommand{\thetable}{A\arabic{table}}
\renewcommand{\theHtable}{appendix.\arabic{table}}
\small
\onecolumn 
\section{Appendix 1 Included studies}\label{appdx:scope}

Table~\ref{apdx:studies} outlines the studies included in this review, as described in Fig.~\ref{fig:prisma} and section~\ref{sec:method}, including details about their research method, sample sizes and geographical region coverage. Data were only coded when explicitly mentioned in the study, otherwise left blank.

\begingroup
\setlength{\tabcolsep}{4pt}
\begin{longtable}{|l|p{0.75cm}|p{2cm}|>{\centering\arraybackslash}p{1.7cm}|>{\raggedright\arraybackslash}p{2.9cm}|>{\raggedright\arraybackslash}p{1.6cm}|p{1cm}|>{\raggedright\arraybackslash}p{1.7cm}|}
\caption{Details of studies included in this review.}\label{apdx:studies}\\
\hline
\textbf{Study} & \textbf{Pub. Year} & \textbf{Study Type} & \textbf{Framework Proposal} & \textbf{Research Method} & \textbf{Data Collection Method} & \textbf{Sample Size} & \textbf{Region} \\
\hline
\cite{Ahmed2020} & 2020 & Conference Paper & - & Qualitative, Case Study & Survey, Interview & - & Pakistan \\ \hline
\cite{Ahmed2021} & 2021 & Journal Article & X & Qualitative & Survey, Literature Review & 200 & Middle East \\ \hline
\cite{Aigbefo2022TheIntention} & 2022 & Journal Article & - & Quantitative & Survey & 294 & USA \\ \hline
\cite{AL-Dosari2023Risk-ManagementApproach} & 2023 & Journal Article & X & Qualitative & Survey, Literature Review & - & - \\ \hline
\cite{Al-Somali2024OrganizationalCulture} & 2024 & Journal Article & X & Qualitative & Survey & 394 & Saudi Arabia \\ \hline
\cite{Alahmari2020} & 2020 & Journal Article & - & Qualitative & Literature Review & - & UK \\ \hline
\cite{Alahmari2021} & 2021 & Conference Paper & - & Qualitative & Interview & 20 & Saudi Arabia \\ \hline
\cite{Alghamdi2019} & 2019 & Conference Paper & X & Literature Review & Literature Review & 40 & - \\ \hline
\cite{Alharbi2021} & 2021 & Journal Article & X & Quantitative & Survey & 282 & Saudi Arabia \\ \hline
\cite{Ali2022CybersecuritySME} & 2022 & Conference Paper & X & Qualitative & Unclear & - & Malaysia \\ \hline
\cite{Aljumaili2018SMEPolicies} & 2018 & PhD Thesis & - & Qualitative & Interview, Observation & 10 & USA \\ \hline
\cite{Almoaigel2023ImplementationSMEs} & 2023 & Journal Article & X & Quantitative & Survey & 350 & Saudi Arabia \\ \hline
\cite{Almubayedh2018} & 2018 & Journal Article & X & Qualitative, Case Study & Interview & 1 & Saudi Arabia \\ \hline
\cite{Asprion2023} & 2023 & Journal Article & X & Qualitative & Interview & 2 & - \\ \hline
\cite{Auyporn2023AStandard} & 2023 & Journal Article & - & Quantitative & Survey & 312 & Thailand \\ \hline
\cite{Azinheira2023} & 2023 & Journal Article & X & Qualitative, Case Study & Survey & 50 & Portugal \\ \hline
\cite{Babalola2021Cybersecurity} & 2021 & PhD Thesis & - & Quantitative & Survey & 187 & USA \\ \hline
\cite{Bada2019} & 2019 & Journal Article & X & Quantitative, Qualitative, Case Study & Survey, Literature Review & 27 & UK \\ \hline
\cite{Bailetti2020} & 2020 & Journal Article & - & Qualitative & - & - & - \\ \hline
\cite{Barlette2019} & 2019 & Journal Article & - & Quantitative, Qualitative & Survey & 200 & - \\ \hline
\cite{Barlette2017} & 2017 & Journal Article & - & Quantitative & Survey & 292 & France \\ \hline
\cite{Berry2018AnThreats} & 2018 & Journal Article & - & Quantitative & Survey, Interview & 373 & USA, Canada \\ \hline
\cite{Borkovich2020} & 2020 & Journal Article & X & Qualitative & Interview & - & USA \\ \hline
\cite{Bountouni2023AStudy} & 2023 & Conference Paper & X & Qualitative, Case Study & - & - & Europe \\ \hline
\cite{Brodin2020MobileApproach} & 2020 & Journal Article & X & Qualitative, Case Study & - & - & Sweden \\ \hline
\cite{Brunner2018} & 2018 & Conference Paper & X & Quantitative, Qualitative, Case Study & Survey, Document Analysis, Observation & 24 & Austria \\ \hline
\cite{Carias2020} & 2020 & Journal Article & X & Qualitative & Interview, Literature Review & 11 & - \\ \hline
\cite{Carias2021CyberSMEs} & 2021 & Conference Paper & X & Qualitative, Case Study & Interview & 11 & - \\ \hline
\cite{Carias2021} & 2021 & Journal Article & X & Qualitative, Case Study & Interview, Literature Review & 3 & El Salvador, Spain, USA \\ \hline
\cite{Cartwright2023CascadingCompanies} & 2023 & Journal Article & X & Qualitative, Case Study & Interview, Literature Review & - & UK \\ \hline
\cite{Chandra2020} & 2020 & Journal Article & X & Qualitative & Survey & 9 & Indonesia \\ \hline
\cite{Chidukwani2022} & 2022 & Journal Article & - & Qualitative & Survey, Literature Review & - & Australia \\ \hline
\cite{Cleveland2018} & 2018 & PhD Thesis & - & Quantitative & Survey & 72 & - \\ \hline
\cite{Cruzado2022Reference27032} & 2022 & Journal Article & X & Quantitative, Case Study & Literature Review, Document Analysis & 1 & Peru \\ \hline
\cite{Dimitrov2019} & 2019 & Journal Article & X & Qualitative & Survey & - & Europe \\ \hline
\cite{Douchek2020} & 2020 & Conference Paper & - & Quantitative & Document Analysis & 440 & Czech Rep., Slovakia \\ \hline
\cite{Durowoju2020} & 2020 & Conference Paper & - & Quantitative, Case Study & - & - & - \\ \hline
\cite{Dykstra2020} & 2020 & Conference Paper & - & Qualitative, Case Study & Survey & 131 & USA \\ \hline
\cite{Diaz-Piraquive2023CybersecurityColombia} & 2023 & Conference Paper & - & Quantitative & Survey & 130 & Colombia \\ \hline
\cite{Ebot2024} & 2024 & PhD Thesis & - & Qualitative & Interview & 12 & - \\ \hline
\cite{Eilts2020} & 2020 & PhD Thesis & - & Quantitative, Qualitative & Survey, Interview & 216 & - \\ \hline
\cite{Eilts2020Article} & 2020 & Journal Article & - & Quantitative, Qualitative & Survey, Interview & 216 & - \\ \hline
\cite{Elezaj2019} & 2019 & Conference Paper & X & Quantitative, Case Study & - & - & Norway \\ \hline
\cite{Erdogan2023} & 2023 & Conference Paper & - & Quantitative & Survey & 141 & UK \\ \hline
\cite{Eybers2022InvestigatingSMEs} & 2022 & Journal Article & - & Qualitative & Interview & 15 & South Africa \\ \hline
\cite{Feher2020} & 2020 & Journal Article & X & Qualitative, Case Study & Interview, Document Analysis, Observation & 2 & Hungary \\ \hline
\cite{Franco2023CyberTEAInvestment} & 2023 & Journal Article & X & Qualitative, Case Study & - & - & Switzerland \\ \hline
\cite{Gafni2019} & 2019 & Journal Article & - & Qualitative & Unclear & - & USA \\ \hline
\cite{Gamisch2023ASecurity} & 2023 & Conference Paper & - & Qualitative & - & - & Europe \\ \hline
\cite{Perez-Gonzalez2019} & 2019 & Journal Article & X & Quantitative & Survey, Interview & 111 & Spain \\ \hline
\cite{VanHaastrecht2021} & 2021 & Journal Article & - & Qualitative & Literature Review & - & Europe \\ \hline
\cite{Heidenreich2019} & 2019 & Conference Paper & X & Quantitative & Literature Review & - & Germany \\ \hline
\cite{Heidenreich2022EvaluationMicro-enterprises} & 2022 & Journal Article & X & Quantitative & Survey, Observation & 30 & Germany \\ \hline
\cite{Heidt2018TheSecurity} & 2018 & Journal Article & X & Qualitative & Interview, Literature Review & 25 & Germany \\ \hline
\cite{Heidt2019} & 2019 & Journal Article & X & Qualitative & Interview, Literature Review & 25 & Western Europe \\ \hline
\cite{Hoong2024NavigatingEnterprises} & 2024 & Journal Article & X & Qualitative & Interview & 35 & Canada \\ \hline
\cite{Huaman2021} & 2021 & Conference Paper & - & Quantitative & Survey & 5000 & Germany \\ \hline
\cite{Harting2023} & 2023 & Journal Article & - & Quantitative, Qualitative & Survey & 104 & Germany \\ \hline
\cite{Ikuero2022} & 2022 & Journal Article & X & Quantitative & Survey, Literature Review & 100 & Nigeria \\ \hline
\cite{Jamil2024Human-centricBehaviours} & 2024 & Journal Article & X & Quantitative & Survey & 502 & Australia \\ \hline
\cite{Jantti2020} & 2020 & Conference Paper & - & Qualitative & Interview & 20 & Finland \\ \hline
\cite{Johannsen2020} & 2020 & Journal Article & X & Case Study & Literature Review & - & Germany \\ \hline
\cite{Felton2021Resilience} & 2021 & PhD Thesis & - & Qualitative & Interview & 5 & USA \\ \hline
\cite{Kabanda2018} & 2018 & Journal Article & - & Qualitative, Case Study & Interview, Document Analysis & 8 & South Africa \\ \hline
\cite{Katt2018} & 2018 & Conference Paper & X & Quantitative, Case Study & - & - & Norway \\ \hline
\cite{Kembumbara2024} & 2024 & PhD Thesis & - & Quantitative & Survey & 183 & USA \\ \hline
\cite{Kljucnikov2019} & 2019 & Journal Article & - & Quantitative & Survey, Literature Review & 10 & Slovakia \\ \hline
\cite{Lejaka2019} & 2019 & Journal Article & - & Qualitative & Literature Review & - & South Africa \\ \hline
\cite{Lejaka2023TowardsSMMEs} & 2023 & Conference Paper & X & Qualitative & Literature Review & - & South Africa \\ \hline
\cite{Lucas2018} & 2018 & PhD Thesis & - & Qualitative & Interview & 10 & USA \\ \hline
\cite{Lynch2020CUI} & 2020 & PhD Thesis & - & Qualitative & Survey & 21 & USA \\ \hline
\cite{Loffler2021} & 2021 & Journal Article & X & Qualitative, Case Study & Interview & - & Switzerland \\ \hline
\cite{MacEwan2017} & 2017 & PhD Thesis & - & Qualitative, Case Study & Interview, Document Analysis, Observation & 3 & UK \\ \hline
\cite{Madison2016Funding} & 2016 & PhD Thesis & - & Qualitative, Case Study & Interview, Document Analysis & 8 & USA \\ \hline
\cite{Marett2019InformationAnalysis} & 2019 & Journal Article & - & Quantitative, Qualitative & Survey, Interview & 232 & USA \\ \hline
\cite{Mayer2017} & 2017 & Journal Article & X & Quantitative, Qualitative, Case Study & Survey & 90 & Germany \\ \hline
\cite{McLaurin2021} & 2021 & PhD Thesis & - & Quantitative, Qualitative & Survey, Document Analysis & 64 & USA \\ \hline
\cite{McLilly2020CloudCybersecurity} & 2020 & PhD Thesis & - & Quantitative & Document Analysis & 564 & USA \\ \hline
\cite{McLilly2020} & 2020 & Journal Article & X & Quantitative, Case Study & - & - & USA \\ \hline
\cite{Mierzwa2017} & 2017 & Technical Report & - & Qualitative & Literature Review & - & - \\ \hline
\cite{Mitrofan2020} & 2020 & Journal Article & - & Quantitative, Case Study & Survey, Interview & 50 & Romania \\ \hline
\cite{Mmango2023CyberICECET} & 2023 & Journal Article & X & Qualitative & Interview, Literature Review & - & - \\ \hline
\cite{Moeti2023InformationEnterprise} & 2023 & Journal Article & X & Qualitative & Interview, Literature Review & 24 & South Africa \\ \hline
\cite{Mokwetli2018} & 2018 & Conference Paper & X & Quantitative & Survey & 647 & South Africa \\ \hline
\cite{Moneva2023} & 2023 & Journal Article & - & Quantitative, Qualitative & Survey & 496 & Netherlands \\ \hline
\cite{Muhammad2023InformationEnterprises} & 2023 & Journal Article & X & Quantitative & Survey & - & Indonesia \\ \hline
\cite{Nasir2020} & 2020 & Conference Paper & - & Qualitative & Interview & 50 & Nigeria \\ \hline
\cite{Ncubukezi2023RiskICCWS} & 2023 & Conference Paper & X & Case Study & Survey, Interview & 25 & South Africa \\ \hline
\cite{Ncubukezi2020} & 2020 & Conference Paper & - & Qualitative & Survey, Interview & 30 & South Africa \\ \hline
\cite{Ncubukezit2022} & 2022 & Journal Article & - & Qualitative & Interview & 30 & South Africa \\ \hline
\cite{Neri2024} & 2024 & Journal Article & - & Quantitative, Qualitative & Survey, Interview & 53 & Italy \\ \hline
\cite{Nkhoma2018InformationSurveys} & 2018 & Journal Article & - & Quantitative, Qualitative & Survey, Interview & 504 & Vietnam \\ \hline
\cite{Ogunyebi2018AnSector} & 2018 & Conference Paper & X & Qualitative, Case Study & Interview & - & - \\ \hline
\cite{Oluwatimilehin2019} & 2019 & PhD Thesis & - & Qualitative & Interview & 10 & USA \\ \hline
\cite{Orellana2020Ecuador} & 2020 & PhD Thesis & - & Qualitative & Survey & 50 & Ecuador \\ \hline
\cite{Oroni2023StructuralPerformance} & 2023 & Journal Article & - & Quantitative & Survey & 97 & Kenya \\ \hline
\cite{Osborn2018} & 2018 & Journal Article & - & Qualitative, Case Study & Survey, Interview & 33 & UK \\ \hline
\cite{Ozkan2019} & 2019 & Conference Paper & X & Qualitative, Case Study & Interview, Literature Review & 20 & Europe \\ \hline
\cite{YigitOzkan2020AddressingModels} & 2020 & Conference Paper & X & Qualitative & - & - & Europe \\ \hline
\cite{Ozkan2023AdaptableSMEs} & 2023 & Journal Article & X & Quantitative, Qualitative & Interview & 7 & Europe \\ \hline
\cite{Ozkan2020} & 2020 & Conference Paper & X & Qualitative, Case Study & - & - & Netherlands \\ \hline
\cite{Park2021} & 2021 & Journal Article & X & Quantitative & Survey & 435 & South Korea \\ \hline
\cite{Pickering2021} & 2021 & Conference Paper & - & Qualitative & Survey & 164 & Europe \\ \hline
\cite{Pickering2023ITechnology} & 2023 & Journal Article & - & Quantitative, Qualitative & Survey, Interview, Focus Group & 800 & UK, Europe \\ \hline
\cite{Ponsard2019SurveySMEs} & 2019 & Journal Article & X & Qualitative & Survey, Interview, Literature Review & - & Belgium \\ \hline
\cite{Ponsard2020MethodologyBelgium} & 2020 & Journal Article & - & Qualitative & Interview & 25 & Belgium \\ \hline
\cite{Pringle2020} & 2020 & PhD Thesis & - & Qualitative & Interview & 10 & - \\ \hline
\cite{Rae2019} & 2019 & Book & X & Quantitative, Qualitative & Survey, Interview & 15 & UK \\ \hline
\cite{Rae2020} & 2020 & Conference Paper & X & Quantitative, Qualitative & Survey & 20 & UK \\ \hline
\cite{Raineri2020} & 2020 & Journal Article & - & Qualitative & - & - & USA \\ \hline
\cite{Rawindaran2021} & 2021 & Journal Article & X & Quantitative, Case Study & Literature Review & - & UK \\ \hline
\cite{Rawindaran2023EnhancingKingdom} & 2023 & Journal Article & X & Quantitative, Qualitative & Survey, Literature Review & 68 & Saudi Arabia, UK \\ \hline
\cite{Rodriguez-Baca2022BusinessSMEs} & 2022 & Journal Article & X & Quantitative, Qualitative, Case Study & Interview, Document Analysis & 2 & Peru, Mexico \\ \hline
\cite{Roy2023FrameworkInCACCT} & 2023 & Journal Article & X & Quantitative & Survey & 125 & - \\ \hline
\cite{Ruhwanya2019InformationTanzania} & 2019 & Journal Article & - & Quantitative, Qualitative & Survey, Focus Group & 34 & Tanzania \\ \hline
\cite{Saa2017} & 2017 & Journal Article & - & Qualitative & - & - & Ecuador \\ \hline
\cite{Saban2021} & 2021 & Journal Article & X & Quantitative & Survey, Interview & - & USA \\ \hline
\cite{Scholl2023SustainableEffect} & 2023 & Journal Article & X & Quantitative, Qualitative & Survey, Interview & 16 & Germany \\ \hline
\cite{Shah2017ISRA} & 2017 & PhD Thesis & - & Qualitative & Survey & 12 & USA \\ \hline
\cite{Shojaifar2020SMEsSharing} & 2020 & Journal Article & X & Qualitative & Interview & 7 & Europe \\ \hline
\cite{Shojaifar2023DesignTool} & 2023 & Journal Article & X & Qualitative & Survey, Interview & 12 & Europe \\ \hline
\cite{Sia2021ReasonsStudies} & 2021 & Journal Article & - & Qualitative, Case Study & Interview, Document Analysis & 2 & Singapore \\ \hline
\cite{Simon2020} & 2020 & Conference Paper & - & Qualitative & Interview & 22 & France \\ \hline
\cite{Tam2021} & 2021 & Conference Paper & - & Qualitative & Literature Review & - & Australia \\ \hline
\cite{Teymourlouei2019} & 2019 & Conference Paper & - & Qualitative & - & - & USA \\ \hline
\cite{Tsimba2021} & 2021 & PhD Thesis & - & Qualitative & Interview, Literature Review & 10 & - \\ \hline
\cite{Tsiodra2023CyberStudy} & 2023 & Journal Article & X & Quantitative, Case Study & Survey & 131 & UK \\ \hline
\cite{Ulrich2020} & 2020 & Conference Paper & - & Quantitative & Survey & 184 & Germany \\ \hline
\cite{White2020} & 2020 & Journal Article & X & Qualitative & Interview & 21 & UK \\ \hline
\cite{Wilson2023} & 2023 & Journal Article & - & Quantitative, Qualitative & Survey & 85 & UK \\ \hline
\cite{Young2020} & 2020 & Conference Paper & X & Qualitative, Case Study & - & - & USA \\ \hline
\cite{Yudhiyati2021} & 2021 & Journal Article & - & Qualitative, Case Study & Interview & 6 & Indonesia \\ \hline
\end{longtable}
\endgroup

\end{appendices}

\bibliographystyle{oup-unsrtnat}
\bibliography{references}



\end{document}